\documentclass[10pt,twocolumn,letterpaper]{article}

\usepackage{graphicx}
\usepackage{makecell}
\usepackage{booktabs,pifont}
\usepackage{amsmath,amsfonts,bm}
\usepackage{algorithm}
\usepackage{algpseudocode}
\newcommand{\cmark}{\ding{51}}
\newcommand{\xmark}{\ding{55}}
\usepackage{multirow}
\usepackage{cvpr}              
\definecolor{cvprblue}{rgb}{0.21,0.49,0.74}
\usepackage[pagebackref,breaklinks,colorlinks,allcolors=cvprblue]{hyperref}

\def\paperID{41635} 
\def\confName{CVPR}
\def\confYear{2026}

\title{Taming Video Models for 3D and 4D Generation via Zero-Shot Camera Control}

\author{
Chenxi Song$^{1}$ \quad 
Yanming Yang$^{1}$ \quad 
Tong Zhao$^{1}$ \quad 
Ruibo Li$^{2}$ \quad 
Chi Zhang$^{1, *}$ \\
\vspace{-0.25cm}\\
$^{1}$AGI Lab, Westlake University \quad 
$^{2}$Nanyang Technological University \quad \\
\vspace{-0.25cm}\\
Project Page: \url{https://worldforge-agi.github.io}
}

\begin{document}
\twocolumn[{
\maketitle
\begin{center}
  \includegraphics[width=\linewidth]{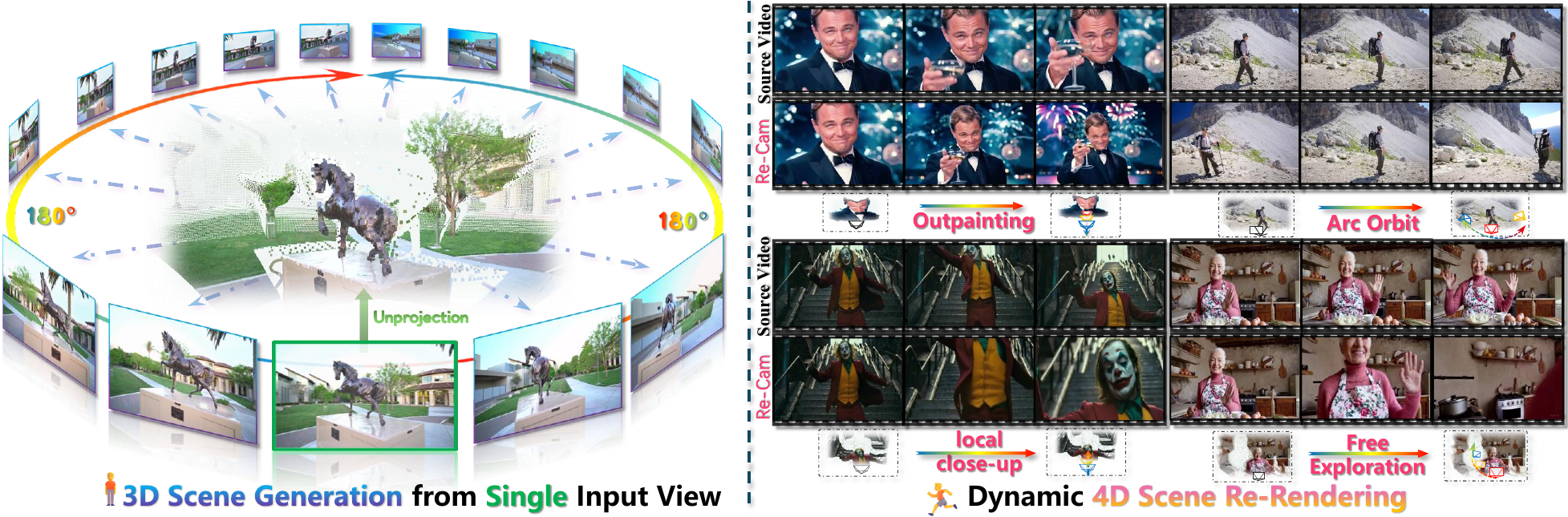}
\end{center}
\captionsetup{type=figure}
\caption{We present \textbf{WorldForge}, a fully training-free framework leveraging a pre-trained video diffusion model for various 3D/4D tasks, such as monocular 3D scene generation (left) and dynamic 4D scene re-rendering (right), enabling precise camera trajectory control and high-quality outputs.}
\label{fig:teaser}}]
\def\thefootnote{*}\footnotetext{Corresponding author.}
\def\thefootnote{\arabic{footnote}}
\addtocontents{toc}{\protect\setcounter{tocdepth}{0}}

\begin{abstract}
Video diffusion models have rich world priors, but their use in spatial tasks is limited by poor control, spatial-temporal inconsistent results, and entangled scene-camera dynamics. Current approaches, such as per-task fine-tuning or post-process warping, often introduce visual artifacts, fail to generalize, or incur high computational costs. We introduce \textbf{WorldForge}, a novel, training-free framework that operates purely at inference time to resolve these issues. Our method comprises three synergistic components. First, an intra-step refinement loop injects fine-grained motion guidance during the denoising process, iteratively correcting the output to ensure strict adherence to the target camera path. Second, an optical flow-based analysis identifies and isolates motion-related channels within the latent space. This allows our framework to selectively apply guidance, thereby decoupling motion from appearance and preserving visual fidelity. Third, a dual-path guidance strategy adaptively corrects for drift by comparing the guided generation against an unguided, reference denoising path, effectively neutralizing artifacts caused by misaligned structural inputs. Together, these components inject precise, trajectory-aligned control without model retraining, achieving accurate motion guidance and photorealistic synthesis. As a plug-and-play, model-agnostic solution, \textbf{WorldForge} demonstrates highly versatile generalizability. Beyond robust zero-shot 3D/4D generation, it readily empowers over a dozen diverse downstream applications, seamlessly enabling tasks like video editing, stabilization, and virtual try-on. Extensive experiments confirm state-of-the-art performance in trajectory adherence and perceptual quality, outperforming both training-dependent and inference-only baselines.
\end{abstract}    
\section{Introduction}
\label{sec:intro}

Recent video diffusion models (VDMs)~\citep{SVD,Wan21,CogVideoX,Veo3} have significantly advanced spatial intelligence \citep{4DSpatialSurvey} tasks like 3D/4D understanding \citep{AC3D,VD3D}, reconstruction \citep{VideoScene,DiFix3DPlus,MVDream}, and generation \citep{ViewCrafter,TrajCrafter}. Trained on vast video datasets, these models encode rich spatiotemporal priors, enabling realistic spatial transformations for applications like novel view synthesis \citep{NVSSolver,TrajectoryAttention}, panoramic video \citep{360DVD,VidPanos}, and dynamic scene reconstruction \citep{ReCamMaster,TrajCrafter,GCD}. Furthermore, VDMs are increasingly used to build ``world models'' \citep{NavigationWM,WorldScore,Genie}, which are structured internal representations that support predictive reasoning in embodied AI.

Despite their strong priors, VDMs face fundamental limitations, including limited controllability, spatial–temporal consistency, and geometric fidelity, particularly when applied to 3D or 4D tasks \citep{MotionCtrl,CameraCtrl,Motionclone,DynamicCrafter}. They struggle to follow precise motion constraints, such as a 6-DoF camera trajectory \citep{AnimateAnyone,FollowYourPose}, which undermines spatial consistency in tasks such as novel view synthesis and trajectory control. These models also entangle scene and camera motion, causing unintended object deformations and scene instability when viewpoint changes are desired \citep{ViewCrafter,ViewExtrapolator}. These limitations hinder their use in applications requiring structured spatial reasoning or controllable generation.

To handle these issues, prior works \citep{ReAngle,Gen3c,TrajCrafter,Recapture} have pursued two main directions. The first, fine-tuning on motion-conditioned data \citep{ReCamMaster,TrajMaster3D,SyncamMaster}, can improve control but is computationally costly, generalizes poorly, and risks degrading pretrained priors. The second, a ``warping-and-repainting" strategy \citep{FYC,Free4D,See3D,NVSSolver}, re-projects frames along a new camera path and uses a generative model to fill occlusions. Although this approach is more flexible, it lacks robustness because pretrained models handle warped, out-of-distribution (OOD) \citep{OOD} inputs poorly, often producing artifacts and fragmented geometry; for example, a bias toward dynamic training data can cause hallucinated motion in static scenes. Consequently, balancing fine-grained controllability with generation quality and generalization remains a challenging open problem.

To address this challenge, we aim to inject precise control into VDMs while preserving their valuable priors. For this purpose, we propose \textbf{WorldForge}, a general inference-time guidance paradigm that leverages the rich priors of VDMs in spatial intelligence tasks, such as geometry-aware 3D scene generation and video trajectory control. Our method uses a warping-and-repainting pipeline, in which input frames are warped along a reference trajectory and then used as conditional inputs in the repainting stage. Building on this, we develop a unified, training-free framework composed of three complementary mechanisms, each designed to address a specific challenge in trajectory-controlled generation, and work synergistically to effectively address the aforementioned OOD challenge.

First, to ensure the generated motion follows the target trajectory derived from depth-based rendering \citep{VGGT,UniDepth}, we introduce Intra-Step Recursive Refinement (IRR). It embeds a micro-scale predict–correct loop within each denoising step: before the next timestep, predicted content in observed regions is replaced with the corresponding ground-truth (GT) observations. This incremental correction allows trajectory control signals to be injected at every step, enabling fine-grained aligned with the target trajectory.

Second, we observe that different channels of the VAE-encoded \citep{VAE, 3DVAE} latent representation encode different information, with some channels specializing in appearance and others in motion. This observation is also noted in \citep{MOFT}, where they remove content correlation and use statistical methods to find channels that contribute most to the principal motion components. Directly overwriting all channels when injecting trajectory signals can inadvertently degrade visual details. To address this, we propose Flow-Gated Latent Fusion (FLF), which leverages optical-flow similarity to selectively inject trajectory information into channels highly correlated with motion, while leaving appearance-relevant channels unmodified. This process effectively decouples appearance from motion, allowing for precise viewpoint manipulation while preserving content fidelity.

Finally, while warping-based rendering effectively enforces user-defined trajectories, it inevitably introduces noise and visual artifacts stemming from imperfect depth, occlusions, or misalignments \citep{NVSSolver}. To balance control with quality, we propose Dual-Path Self-Corrective Guidance (DSG). Inspired by CFG \citep{CFG}, DSG uses two parallel denoising paths during inference: A non-guided path that relies on the model's priors to produce high-fidelity but uncontrolled results, and a guided path that follows the warped trajectory, ensuring camera control but risking artifacts. At each step, DSG computes the difference between these paths to create a dynamic correction term. This term adjusts the guided path toward the higher perceptual quality of the non-guided path. This self-corrective mechanism mitigates artifacts from the warped trajectory while maintaining camera control, improving the video's overall structure and visual quality.

Together, these three mechanisms form a cohesive inference-time guidance framework for robust and precise trajectory control while preserving VDM priors. Our method is training-free and plug-and-play, enabling broad applicability across tasks without model retraining. It is also model-agnostic and readily adapts to backbones such as Wan 2.1 \citep{Wan21} and SVD \citep{SVD}. Comprehensive experiments on multiple tasks and benchmarks confirm that our approach achieves state-of-the-art (SOTA) results, improving trajectory adherence, geometric consistency, and perceptual quality over leading baselines.
Our main contributions are:
\begin{itemize}
  \item\textbf{WorldForge}, a novel, training-free paradigm for leveraging VDM priors in spatial intelligence tasks, enabling precise and stable 3D/4D trajectory control without retraining or fine-tuning.
  \item A synergistic inference-time guidance framework integrating Intra-Step Recursive Refinement (IRR) and Flow-Gated Latent Fusion (FLF), achieving accurate trajectory adherence while decoupling motion from content.
  \item Dual-Path Self-Corrective Guidance (DSG), a self-referential correction mechanism that enhances spatial alignment and perceptual fidelity without auxiliary networks or retraining.
  \item Extensive experiments on diverse datasets and tasks show our approach achieves SOTA controllability and visual quality, even compared to training-intensive pipelines.
  \vspace{-0.5em}
\end{itemize}
\begin{figure*}[!t]
    \centering
    \includegraphics[width=1\linewidth]{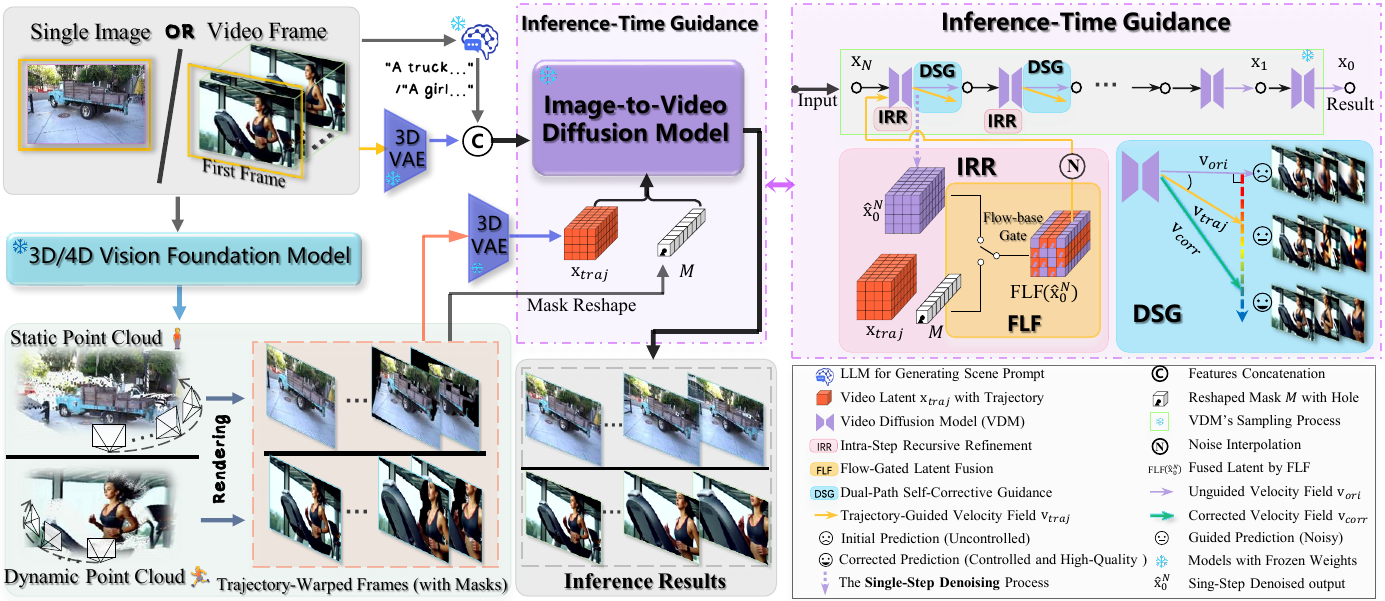}
    \vspace{-1em}
    \caption{Overview of \textbf{WorldForge}. Given a single image or video frames, a vision foundation model reconstructs a scene point cloud, which is warped and rendered along a user-specified trajectory to produce a guidance video. The input image (or first frame) is also converted into a textual prompt and latent representation for an image-to-video diffusion model. Trajectory control is injected through a training-free strategy comprising IRR, FLF, and DSG (detailed in Sec.~3.2–3.4), enabling precise control and high-quality synthesis without additional training.}
    \vspace{-0.5em}
    \label{fig:pipeline}
\end{figure*}

\section{Related Works}
\label{sec:related_works}

We review prior work in three relevant areas: 3D static scene generation, 4D trajectory-controlled video generation, and guidance strategies for generative models.

\noindent\textbf{3D Static Scene Generation.} 
While 3D reconstruction \citep{NeRF,3DGS,Fewarnet,CAT3D,Mipsplatting,InstantNGP,MVSNet} and object generation \citep{DreamFusion,ASD,Trellis,GeometryAwareSDS} are advanced, they often lack scene-level priors. VDM \citep{SVD, Wan21, HunyuanVideo} provide these priors and are leveraged by decoding scenes from images \citep{Wonderland}, fine-tuning on warped inputs \citep{ViewCrafter, See3D}, or embedding camera parameters \citep{MotionCtrl, TrajectoryAttention}. Unlike costly fine-tuning which can corrupt priors, training-free strategies \citep{NVSSolver, ViewExtrapolator} preserve them but must ensure geometric coherence. Our work takes this training-free approach to enhance both consistency and control.

\noindent\textbf{Trajectory-Controlled Dynamic Video Generation.} One paradigm for controllable video synthesis is to fine-tune lightweight adapters \citep{FYC, T2IAdapter,4Real,MotionCtrl} like LoRA \citep{LoRA} or ControlNet \citep{ControlNet} on video-trajectory data, conditioning on diverse inputs \citep{ReCamMaster,TrajCrafter,GCD,DaS}. Another is the \emph{warp-and-repaint} strategy \citep{FYC, Free4D,Voyager, VUD}, which projects and inpaints frames but is prone to artifacts from noisy warps \citep{NVSSolver}. Our work uses inference-time guidance to directly steer the diffusion process for precise motion control.

\noindent\textbf{Guidance and Control for Generative Models.} Guidance strategies steer diffusion models toward desired outputs. While Classifier-Free Guidance (CFG) \citep{CFG} is common, high weights can cause artifacts. More advanced techniques use auxiliary models \citep{BadGuide,STG,RestartSampling} or iterative refinement \citep{zigzag} to improve sampling. In 3D/4D synthesis, guidance is used to enforce viewpoint consistency, but warp-based methods often suffer from noise-induced artifacts \citep{D2T,VistaDream}. To address this, we propose DSG. It derives a correction signal from the difference between guided and unguided predictions at each step, enhancing trajectory adherence and stability without retraining.
\section{Proposed Methods}
\label{sec:methods}

We propose an inference-time guidance strategy to balance controllability with visual fidelity for VDMs in 3D/4D tasks. Our method is a training-free framework that steers a pretrained model along a user-defined trajectory while preserving its generative priors.
As shown in Fig.~\ref{fig:pipeline}, our framework has three key components.
First, Intra-Step Recursive Refinement (IRR) injects trajectory guidance from observed regions at each denoising step for consistent control (Sec.~\ref{sec:3.2}).
Second, Flow-Gated Latent Fusion (FLF) decouples motion from appearance in the latent space to prevent content drift and preserve fidelity (Sec.~\ref{sec:3.3}).
Finally, Dual-Path Self-Corrective Guidance (DSG) uses the difference between guided and unguided paths as a corrective signal to suppress artifacts and improve stability (Sec.~\ref{sec:3.4}).
Together, these modules enable fine-grained trajectory control and unlock the model's latent 3D/4D awareness without any retraining.

\subsection{Preliminaries}

Before detailing our method, we introduce the necessary  preliminaries: diffusion models, guidance strategies, and trajectory-controlled video synthesis.

\subsubsection{Denoising Diffusion Models and Guidance}
\label{sec:3.1.1}
\textbf{Diffusion Solvers.} Generative models are largely diffusion~\citep{DDPM} or flow-based~\citep{Flow}. Under the SDE view, diffusion models have a deterministic ODE limit that connects to flow models via reparameterization~\citep{DiffusionFlow} (The detailed derivation is provided in Supplementary). We use the popular DDIM sampler \citep{DDIM} as an example to illustrate the sampling process: it recovers the clean sample $\mathbf{x}_0$ by reversing the forward noising of a Gaussian prior $\mathbf{x}_T$. Given a noise-prediction network $\bm{\epsilon}_\theta(\mathbf{x}_t,t)$, the sampler estimates an intermediate signal $\hat{\mathbf{x}}_0$ from the current state $\mathbf{x}_t$:
\begin{equation}
\label{eq:x0}
\hat{\mathbf{x}}_0(\mathbf{x}_t, t) = \frac{\mathbf{x}_t - \sqrt{1-\bar{\alpha}_t}\,\bm{\epsilon}_\theta(\mathbf{x}_t, t)}{\sqrt{\bar{\alpha}_t}},
\end{equation}
where $\bar{\alpha}_t$ denotes cumulative noise attenuation. The term $\hat{\mathbf{x}}_0(\mathbf{x}_t, t)$ is a key intermediate variable: at each step, it is the one-step denoised estimate from $\bm{\epsilon}_\theta$, evolving from a coarse prediction to a sharp final output. The next sample $\mathbf{x}_{t-1}$ is then obtained by blending $\hat{\mathbf{x}}_0$ with the predicted noise $\bm{\epsilon}_\theta$:
\begin{equation}
\label{eq:xt-1}
\mathbf{x}_{t-1} = \sqrt{\bar{\alpha}_{t-1}}\,\hat{\mathbf{x}}_0(\mathbf{x}_t, t) + \sqrt{1-\bar{\alpha}_{t-1}}\,\bm{\epsilon}_\theta(\mathbf{x}_t, t).
\end{equation}

Iterating from $t=T$ to $t=0$ yields the final sample $\mathbf{x}_0$. Our method intervenes at this stage by \emph{modifying $\hat{\mathbf{x}}_0$ to enforce trajectory control}. Notably, other popular solvers, such as UniPC \citep{UniPC}, EDM \citep{EDM}, and PNDM \citep{IPNDM}, also compute $\hat{\mathbf{x}}_0$ directly or can recover it via a parameterized transformation, so our framework is broadly compatible and can be flexibly applied to current mainstream diffusion-based and flow-based models. Since our experiments primarily use the flow-based Wan2.1 \citep{Wan21} model, we will subsequently detail our algorithm using a \emph{flow-based} formulation.

\noindent\textbf{Classifier Free Guidance.} To improve fidelity to the condition, CFG \citep{CFG} adjusts the network prediction during sampling as:
\begin{equation}
\label{eq:cfg}
\tilde{\bm{\epsilon}}_\theta(\mathbf{x}_t, t) = \bm{\epsilon}_\theta(\mathbf{x}_t, t, \mathbf{c}) + \omega_\text{CFG} \cdot \left[\bm{\epsilon}_\theta(\mathbf{x}_t, t, \mathbf{c}) - \bm{\epsilon}_\theta(\mathbf{x}_t, t,\phi)\right],
\end{equation}
where $\omega_\text{CFG}$ is the guidance weight, with $\mathbf{c}$ and $\phi$ denoting the conditional and unconditional inputs, respectively. This interpolates conditional and unconditional scores to steer the sampling trajectory. Additionally, other works, such as APG, have sought to address issues like color oversaturation by adjusting the scaling of the guidance term. Our approach extends this principle through a self-referential guidance mechanism that dynamically adjusts the guided prediction using the model’s own unguided output at each step.

\subsubsection{Trajectory Control via Depth-Based Warping}
Our framework controls trajectories using a depth-based warping-and-repainting strategy. First, a depth-prediction network estimates camera poses and depth maps $(\mathbf{P}_q, \mathbf{D}_q)$ from single image ${\mathbf{I}}$ or image sequence $\{\mathbf{I}_q\}_{q=1}^N$ via a function $f:\{\mathbf{I}_q\}_{q=1}^N \rightarrow \{\mathbf{P}_q,\mathbf{D}_q\}$. 
Next, a warping operator $\mathcal{W}$ uses these estimates to project a source frame $\mathbf{I}_{src}$ with depth $\mathbf{D}_{src}$ from pose $\mathbf{P}_{src}$ to a target pose $\mathbf{P}_{tar}$. This yields a partial target view $\mathbf{I}'_{tar}$ and a validity mask $\mathbf{M}_{tar}$ indicating visible pixels:
\begin{equation}
\label{eq:M}
(\mathbf{I}'_{tar}, \mathbf{M}_{tar}) = \mathcal{W}(\mathbf{I}_{src}, \mathbf{D}_{src}, \mathbf{P}_{src}, \mathbf{P}_{tar}).
\end{equation}

The resulting warped frames and masks guide the VDMs along the target poses $\mathbf{P}_{tar}$. This guidance is limited to regions visible in the source views. With these preliminaries, we use trajectory control to guide video generation.
\subsection{Intra-Step Recursive Refinement}
\label{sec:3.2}
To enable precise trajectory injection during VDM's inference processing, we introduce Intra-Step Recursive Refinement (IRR). As noted in Sec.~\ref{sec:3.1.1}, the denoising process produces an intermediate variable $\hat{\mathbf{x}}_0^{(t)}$, a coarse estimate of the final output and the baseline for later steps,where $t$ denotes the current timestep. IRR modifies $\hat{\mathbf{x}}_0^{(t)}$ to impose trajectory constraints, ensuring that generation follows the desired path.

IRR operates within the updates of Eq.~(\ref{eq:x0}) and Eq.~(\ref{eq:xt-1}). Given the one-step denoised sample $\hat{\mathbf{x}}_0^{(t)}$ from Eq.~(\ref{eq:x0}), we fuse it with the trajectory latent $\mathbf{x}_{\text{traj}}$, obtained by VAE-encoding the warped frames $\mathbf{I}'_{tar}$ of Eq.~(\ref{eq:M}) into latent space. We then add Gaussian noise $\bm{\epsilon}$ to obtain the modified latent $\mathbf{x}_{t}'$:
\begin{equation} 
\label{eq:t'}
\mathbf{x}_{t}' = (1 - w (\sigma))\,\mathbf{F}(\hat{\mathbf{x}}_0^{(t)},\mathbf{x}_{\text{traj}}) + w (\sigma)\, \cdot \bm{\epsilon},
\end{equation}
where $\mathbf{F}(\hat{\mathbf{x}}_0^{(t)},\mathbf{x}_{\text{traj}})=\mathbf{M}\cdot \mathbf{x}_{\text{traj}}+(1-\mathbf{M})\cdot \hat{\mathbf{x}}_0^{(t)}$ copies observable warped content from $\mathbf{x}_{\text{traj}}$ into the corresponding locations of $\hat{\mathbf{x}}_0^{(t)}$ using the binary mask $\mathbf{M}$ from Eq.~(\ref{eq:M}). This clean-space fusion on $\hat{\mathbf{x}}_0^{(t)}$, which differs from prior inpainting works like \citet{Repaint} that operate in the noisy $\mathbf{x}_{t-1}$ space, is critical for our subsequent FLF module. 
$\bm{\epsilon}=\mathbf{x}_{T}\sim\mathcal{N}(\mathbf{0},\mathbf{I})$ is a randomly sampled Gaussian noise used to re-noise the fused latent $\mathbf{F}(\hat{\mathbf{x}}_{0}^{(t)},\mathbf{x}_{\text{traj}})$ so that it re-enters the denoising schedule with the injected trajectory. The re-noising is controlled by a scheduler $w(\sigma)$, and the specific strategy can differ for various diffusion and flow models. The re-noised latent $\mathbf{x}_{t}'$ is then fed into the network $\bm{\epsilon}_{\theta}$, replacing the original $\mathbf{x}_{t}$ in Eq.~(\ref{eq:x0}) and Eq.~(\ref{eq:xt-1}) for the next sampling step. In summary, IRR embeds a micro predictor–corrector at each denoising step. By updating $\hat{\mathbf{x}}_0^{(t)}$ with explicit trajectory cues $\mathbf{x}_{\text{traj}}$, it continually corrects the sampling path and ensures that synthesis follows the target trajectory precisely.

\subsection{Flow-Gated Latent Fusion}
\label{sec:3.3}

In the IRR process, overwriting all latent channels with trajectory information degrades visual quality because VAE latents encode both motion and appearance.This observation is supported by \citet{MOFT}, which used PCA to statistically identify channels with distinct motion properties. The indiscriminate update in Eq.~(\ref{eq:t'}) injects noise into appearance-focused channels. To address this, we propose Flow-Gated Latent Fusion (FLF), a method that identifies and updates latent channels with high motion relevance. Unlike \citet{MOFT}, we use an optical-flow-based score that directly describes motion to filter channels, requiring no gradient-based optimization.

To select motion-related channels, we use an optical-flow-based scoring scheme since flow directly reflects inter-frame motion. First, for each channel $c$ of the latent $\hat{\mathbf{x}}_0^{(t)}$ at timestep $t$ (the $\hat{\mathbf{x}}_0^{(t)}$ denotes the one-step denoised prediction at timestep $t$), we compute the optical flow between consecutive frames to get a predicted flow $\mathcal{F}^{(t,c)}_{\text{pred}}$. The resulting map for each channel has a shape of $[2, \tau, H, W]$ (flow vectors, frame pairs, spatial dimensions). Second, we compute a GT reference flow $\mathcal{F}^{(t,c)}_{\text{gt}}$ from the target trajectory latent $\mathbf{x}_{\text{traj}}$ in the same manner. Finally, we compare the two flows within the visible regions defined by the mask $\mathbf{M}^{(c)}$.

The comparison relies on three popular optical flow metrics \citep{RAFT}: Masked End-point Error (M-EPE), which measures the Euclidean distance between predicted and GT flow vectors; Masked Angular Error (M-AE), which measures their angular difference; and Outlier Percentage (Fl-all), which calculates the fraction of unreliable pixels. We combine the normalized metrics to calculate a motion similarity score $S^{(t,c)}$ for each channel in each timestep $t$: 
\begin{equation}
\setlength{\abovedisplayskip}{4pt}
\setlength{\abovedisplayshortskip}{4pt}
\setlength{\belowdisplayskip}{4pt}
\setlength{\belowdisplayshortskip}{4pt}
S^{(t,c)}=\sum_{k\in\{\text{E, A, F}\}} \gamma_k\!\left(1 - {\text{Norm}_k}^{(t,c)}\right),
\label{eq:score}
\end{equation}
where $\text{Norm}_k^{(t,c)}\!\in[0,1]$ are the normalized errors from M-EPE, M-AE, and Fl-all, and $\gamma_k$ are the weighting factors. The calculation and normalization methods for these metrics are provided in the Supplementary. Higher $S^{(t,c)}$ means better flow alignment and stronger motion evidence.

To adaptively set the motion similarity threshold for each scene, we select motion-relevant channels using a dynamic threshold $\delta^{(t)}=\mu_S^{(t)}-\lambda^{(t)}\sigma_S^{(t)}$, where $\mu_S^{(t)}$ and $\sigma_S^{(t)}$ are the mean and standard deviation of all channel scores at step $t$. We schedule $\lambda^{(t)}$ to create a loose-to-tight selection over time. This matches the generative process, where early steps define broad structures and later steps handle fine details. Consequently, we disable flow-gating in the early steps (e.g., the first 5) to use all channels for structural integrity, then progressively reduce the number of guided channels in later steps to preserve high-fidelity details. Specific schedule settings are detailed in the Supplementary.

Finally, the latent update selectively fuses the trajectory latent $\mathbf{x}_{\text{traj}}$ into the selected high motion-relevance channels:
 \begin{equation}
\label{eq:flf_update}
\mbox{\small
$ \displaystyle 
\mathbf{FLF}(\hat{\mathbf{x}}_0^{(t)}\!,\!\mathbf{x}_{\text{traj}})\!\!=\!\!
\begin{cases}
\mathbf{M}^{(c)}\mathbf{x}_{\text{traj}}^{(c)} \!+\! \left(1\! \!-\! \!\mathbf{M}^{(c)}\right) \!\hat{\mathbf{x}}_0^{(t,c)},\!\!\! & \text{if}\; S^{(t, c)} \!\ge\! \delta^{(t)}\\
\hat{\mathbf{x}}_0^{(t,c)}, & \text{otherwise}.
\end{cases}
$
}
\end{equation}

This FLF operator replaces the operator $\mathbf{F}$ in Eq.~(\ref{eq:t'}), resulting in a more precise fusion rule.

\begin{figure}[!t]
    \centering
\includegraphics[width=1\linewidth]{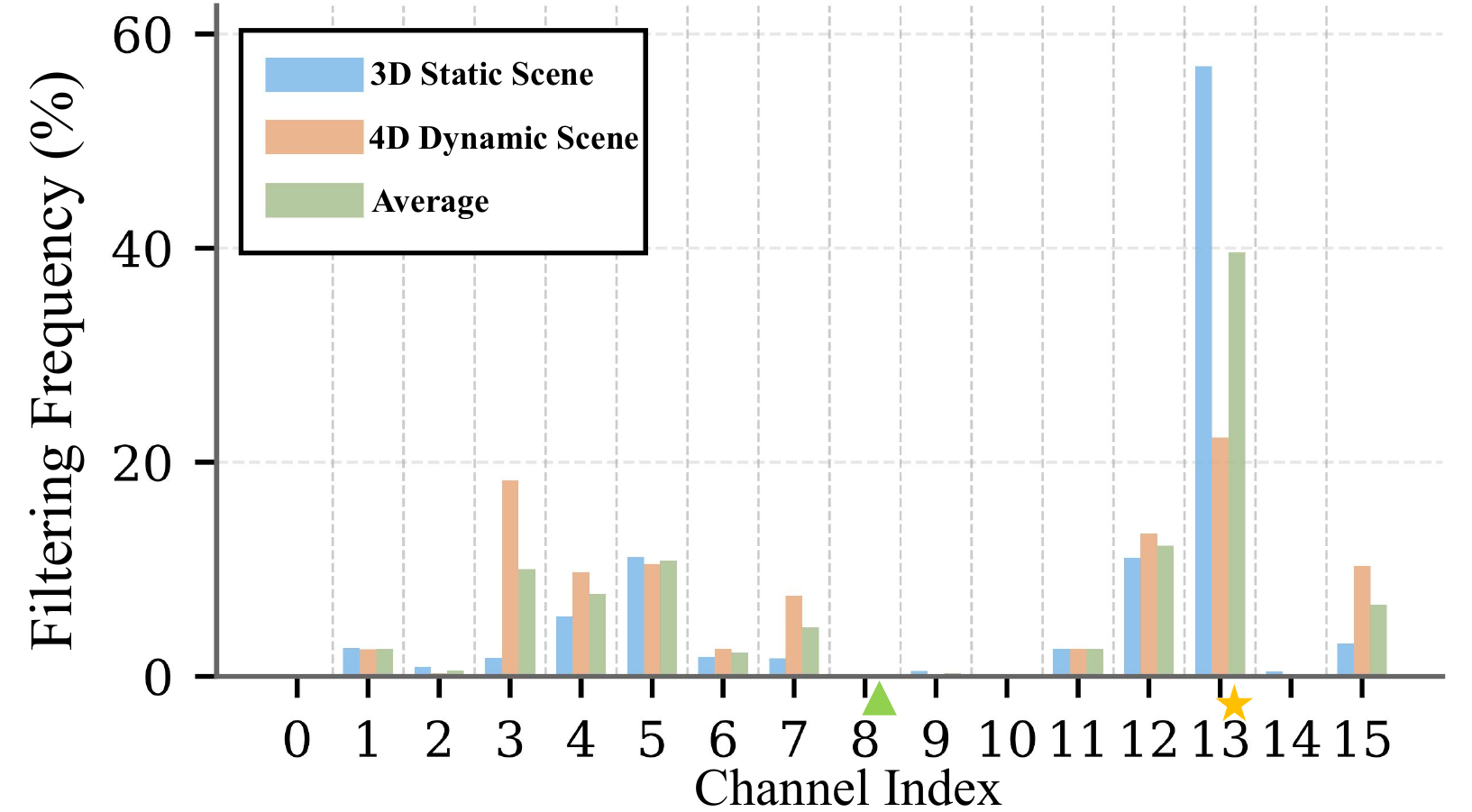}
    \caption{FLF channel-wise flow statistics. The Y-axis shows filtering frequency; a ``filtered out'' channel has a low optical flow score (poor motion correlation). Statistics were gathered by tracking indices at each step across 40+ static and 30+ dynamic scenes. The results confirm distinct, stable roles: channel 13 is most frequently filtered out (low motion relevance), while channel 8 is never filtered out (high motion relevance). 4D dynamic scenes show more diverse scores than 3D scenes, reflecting greater motion complexity. This validates our selective guidance approach.}
    \vspace{-0.5em}
    \label{fig:flf}
\end{figure}

In summary, FLF provides fine-grained trajectory control while preserving model priors and synthesis quality. Unlike methods that restart the sampling schedule \citep{RestartSampling},  iterate in noisy space \citep{Repaint}, or globally update the entire latent \citep{ViewExtrapolator}, FLF integrates with our IRR framework to apply selective, per-step guidance, effectively decoupling motion and appearance for precise control. Fig.~\ref{fig:flf} shows the statistics based on optical flow, demonstrating that each channel exhibits different motion relevance.

\subsection{Dual-Path Self-Corrective Guidance}
\label{sec:3.4}
Trajectory latents $\mathbf{x}_{\text{traj}}$ obtained by warping often contain distortions from depth errors, occlusions, or misalignments, which degrades synthesis quality. To mitigate this, we draw inspiration from CFG~\citep{CFG}. Conceptually, in the context of a flow model, CFG treats the unconditional prediction as a ``bad'' direction $\mathbf{v}_{\text{uncon}}$ and the conditional one as a ``good'' direction $\mathbf{v}_{\text{con}}$ \citep{BadGuide}. It then finds a ``better'' direction by pushing the ``good'' one away from the ``bad'' one. Based on this idea, we propose Dual-Path Self-Corrective Guidance (DSG). At each iteration, IRR produces two velocity fields. The unguided velocity $\mathbf{v}_{t}^{\text{ori}}$ (from the original latent $\mathbf{x}_{t}$) stays on the data manifold with high fidelity but ignores the trajectory, which we consider a ``bad'' direction for control. The guided velocity $\mathbf{v}_{t}^{\text{traj}}$ (from the corrected latent $\mathbf{x}'_{t}$) may be noisy but follows the trajectory, which we consider a ``good'' direction. DSG uses the difference between them to find a ``better'' path.

\begin{figure*}[!t]
    \centering
    \includegraphics[width=1\linewidth]{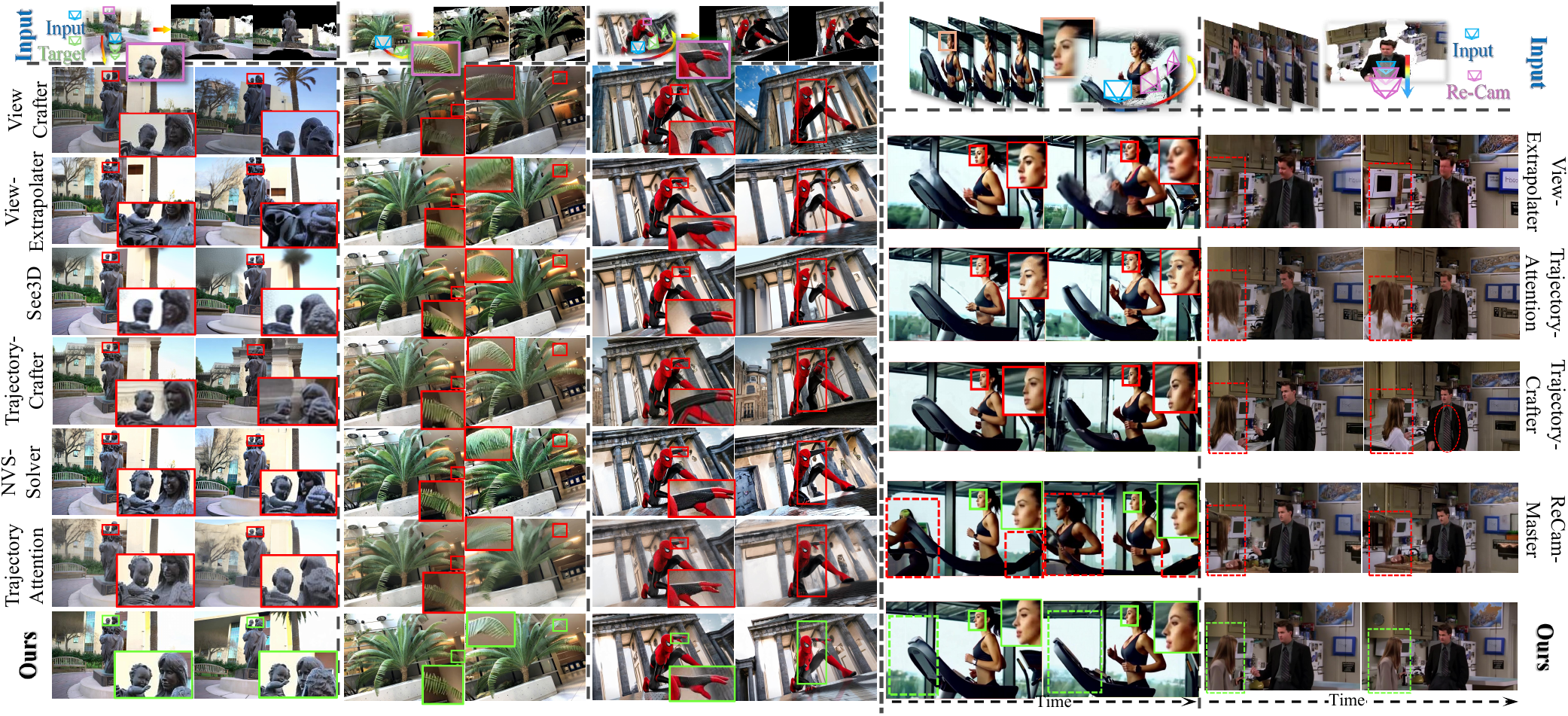}
    \vspace{-1em}
    \caption{Qualitative comparison for 3D novel view synthesis and 4D trajectory-controlled re-rendering. (\textbf{Left}) For 3D scene generation from a single image, our method produces more consistent and plausible content compared to SOTA baselines. (\textbf{Right}) For 4D re-rendering, our approach leverages model priors to avoid common artifacts like floating heads and flattened faces, yielding more realistic results. Overall, our training-free guidance demonstrates superior performance in both static and dynamic scenes. Zoom in for more details.}
\label{fig:3d&4d}
\end{figure*}
In essence, DSG robustly balances precise trajectory control with high visual fidelity. Its adaptive cosine-weighting is crucial for handling the large angular difference between the guided and unguided paths. It suppresses artifacts by applying strong corrections when the paths diverge, while preserving the model's natural predictions when they align. This ensures the final corrected velocity, $\mathbf{v}_{t}^{\text{corr}}$, steers the sample along the target motion path while maintaining the quality of the model's priors.

However, we empirically find the difference between our $\mathbf{v}_{t}^{\text{traj}}$ and $\mathbf{v}_{t}^{\text{ori}}$ is far greater than that between the $\mathbf{v}_{\text{con}}$ and $\mathbf{v}_{\text{uncon}}$ in CFG. In extensive tests, we observed that the cosine similarity $\alpha_{t} = (\mathbf{v}_{t}^{\text{traj}} \cdot \mathbf{v}_{t}^{\text{ori}}) / (\lVert \mathbf{v}_{t}^{\text{traj}} \rVert \cdot \lVert \mathbf{v}_{t}^{\text{ori}} \rVert)$ between our two paths is typically between 0.3--0.6 (an angle of 50°--70°). In contrast, the cosine similarity between $\mathbf{v}_{\text{con}}$ and $\mathbf{v}_{\text{uncon}}$ is nearly 1 (an angle close to 0°). This discrepancy arises because CFG inputs differ only by a prompt. Our trajectory guidance, however, significantly updates the network input $\mathbf{x}_{t}'$. This leads to a large directional divergence between the guided ($\mathbf{v}_{t}^{\text{traj}}$) and unguided ($\mathbf{v}_{t}^{\text{ori}}$) velocity fields. Therefore, directly applying the CFG fails in our case. To address this large angular difference, we modify the guidance formula to only use the component of the ``good'' direction that is orthogonal to the ``bad'' direction. This is achieved by projecting $\mathbf{v}_{t}^{\text{traj}}$ onto $\mathbf{v}_{t}^{\text{ori}}$ (after normalizing them to equal length) and taking the difference, which avoids the adverse effects of their large directional divergence, This principle of isolating and using the orthogonal guidance component was also explored in APG~\citep{APG} to eliminate oversaturation. However, the task and input settings we address are completely different. We extend this principle to our trajectory guidance task, which suffers from a much larger velocity fields angular divergence than APG:
\begin{equation}
\mathbf{v}_{t}^{\text{corr}} = \mathbf{v}_{t}^{\text{traj}} + \rho \cdot \beta_{t} \bigl( \mathbf{v}_{t}^{\text{traj}} - \alpha_{t} \cdot \mathbf{v}_{t}^{\text{ori}} \bigr),
\end{equation}
where $\rho$ controls guidance strength, $\alpha_{t}$ is the cosine similarity, and $\beta_{t} = \sqrt{1-\alpha_{t}^2}$ is its sine counterpart. The role of $\beta_{t}$ is to adaptively scale the guidance: it amplifies the correction when the paths diverge (low $\alpha_{t}$, high $\beta_{t}$) and reduces it when they agree, preserving the model's natural prediction.

\section{Experiments}
\label{sec:experiments}

In this section, we present a comprehensive evaluation of our proposed training-free framework. We first outline the implementation details in Sec.~\ref{sec:4.1}. Subsequently, we demonstrate the performance of our method on 3D scene generation and 4D trajectory control in Sec.~\ref{sec:4.3}. Finally, we conduct a series of ablation studies in Sec.~\ref{sec:4.4} to validate the effectiveness of each component of our approach.

\subsection{Implementation Details}
\label{sec:4.1}
Our framework is a training-free method that steers pre-trained VDMs for precise camera control, with no additional training or fine-tuning. At inference time, our IRR is applied during approximately the first 35-45\% of the steps. For example, when using the Wan2.1 \citep{Wan21} model with a UniPC \cite{UniPC} sampler to generate a video in 50 steps, our IRR is applied during approximately the first 20 steps, but users can fine-tune this number to 15-25 based on the scene for optimal results. 

\noindent\textbf{Setup.} Experiments primarily use the Wan2.1 Image-to-Video (I2V-14B) model~\citep{Wan21}. Generation runs on a single GPU with $\ge$69\,GB VRAM, producing videos up to 1280$\times$720. The per-pass sequence length depends on the chosen VDM’s capacity; wider-view videos are obtained by concatenation. We also evaluate on SVD~\citep{SVD}, which runs on a 24\,GB GPU for 25-frame inference, with qualitative results shown in Fig.~\ref{fig:vdmablation}. However, it is worth noting that SVD's implicit world knowledge is limited by its parameter count and training data, which restricts our method's performance. To further validate the transferability of our method, we also applied it to the recent LongCat-Video \citep{Longcat} model, again achieving results that surpass SOTA methods.

\noindent\textbf{Test Datasets and Metrics.}
For single-view 3D scene generation, we use data from LLFF~\citep{LLFF}, Tanks and Temples~\citep{Tanks}, MipNeRF~360~\citep{Mipnerf360}, and diverse images such as portraits and AI-generated images (e.g., from Pixabay) to validate generalization, testing on 70+ single views from 40+ scenes for quantitative comparison. We report FID~\citep{FID} and $\mathrm{CLIP}_{\mathrm{sim}}$~\citep{CLIP}. As our single-view task does not have GT views, because it is a generation task rather than a reconstruction task, reference-based metrics like PSNR cannot measure the quality of ``imagined'' unseen scenes \citep{TrajectoryAttention, NVSSolver}. For 4D trajectory control, we test on 50+ videos generated from 30+ selected videos using different camera paths (sourced from DAVIS \citep{DAVIS}, movie clips, and VDM generations), reporting FVD~\citep{FVD} and $\mathrm{CLIP\text{-}V}_{\mathrm{sim}}$. Trajectory accuracy for both tasks is measured via ATE, RPE-T, and RPE-R, which are standard metrics for trajectory and geometric consistency \citep{FYC, CameraCtrl}.

\subsection{3D and 4D Trajectory-Controlled Generation}
\label{sec:4.3}
We compare our method against state-of-the-art baselines on both 3D static scene generation and 4D dynamic video control. For 3D novel view synthesis, we evaluate against both training-based \citep{ViewCrafter, TrajCrafter, TrajectoryAttention, See3D} and training-free \citep{NVSSolver, ViewExtrapolator} methods. For 4D trajectory control, baselines include ReCamMaster \citep{ReCamMaster} and others \citep{TrajCrafter, TrajectoryAttention, ViewExtrapolator}. Since ReCamMaster \citep{ReCamMaster} uses a Text-to-Video model, it cannot accept the same video-depth-based warped input to ensure an identical trajectory as the other methods. Therefore, to ensure a fair comparison, we only compare its qualitative visual results.

As shown in Fig.~\ref{fig:3d&4d}, Table~\ref{tab:static}, and Table~\ref{tab:dynamic}, our training-free method consistently achieves superior results under identical evaluation settings. On 3D static scenes, it outperforms both training-based and training-free baselines on public datasets. On 4D clips with diverse camera paths (e.g., arcs, dolly zooms), it yields higher visual fidelity, tighter trajectory alignment, and more coherent scene completion, matching or surpassing costly training-based approaches. In both settings, our method plausibly reconstructs unseen regions where baselines often produce distortions.

Our approach particularly excels in difficult cases, handling human-centric scenes that require high consistency and synthesizing photorealistic $360^\circ$ views from a single input. By preserving model priors, it strikes a strong balance between controllability and fidelity. Furthermore, it serves as a versatile post-production tool for video editing tasks like object addition, removal, replacement, and virtual try-on. Additional visual results and detailed discussions on computational efficiency, robustness across flow estimators, and failure cases are provided in the Supplementary.

\begin{table}[!t]
\centering
\scriptsize
\setlength{\tabcolsep}{1pt} %
\renewcommand{\arraystretch}{1}
\begin{tabular}{l cc ccc c} 
\toprule
& \multicolumn{2}{c}{\textbf{Generation Quality}} & \multicolumn{3}{c}{\textbf{Trajectory Accuracy}} & \multirow{2}{*}{\textbf{\makecell[c]{Training \\ -Free}}} \\
\cmidrule(lr){2-3}\cmidrule(lr){4-6} 
& FID $\downarrow$ & $\mathrm{CLIP}_{\mathrm{sim}} \uparrow$ 
& ATE $\downarrow$ & RPE-T $\downarrow$ & RPE-R $\downarrow$ & \\
\midrule
See3D \citep{See3D} 
& 123.26 & \underline{0.941}
& 0.091 & \underline{0.089} & \underline{0.250}
& \textcolor{red}{\ding{55}} \\ 
ViewCrafter \citep{ViewCrafter} 
& 117.50 & 0.930
& 0.236 & 0.315 & 0.728
& \textcolor{red}{\ding{55}} \\ 
ViewExtrapolator \citep{ViewExtrapolator} 
& 125.50 & 0.930
& 0.183 & 0.260 & 0.882
& \textcolor{green}{\ding{51}} \\ 
TrajectoryAttention \citep{TrajectoryAttention} 
& 122.37 & 0.920
& 0.159 & 0.238 & 0.532
& \textcolor{red}{\ding{55}} \\ 
TrajectoryCrafter \citep{TrajCrafter} 
& \underline{111.49} & 0.910
& \underline{0.090} & 0.152 & 0.267
& \textcolor{red}{\ding{55}} \\ 
NVS\mbox{-}Solver \citep{NVSSolver} 
& 118.64 & 0.937
& 0.224 & 0.268 & 1.056
& \textcolor{green}{\ding{51}} \\ 
\midrule
\textbf{WorldForge (Ours)} 
& \textbf{96.08} & \textbf{0.948}
& \textbf{0.077} & \textbf{0.086} & \textbf{0.221}
& \textcolor{green}{\ding{51}} \\ 
\bottomrule
\end{tabular}
\caption{Quantitative comparison with existing methods on 3D static scenes. We evaluate generation quality (FID, $\mathrm{CLIP}_{\mathrm{sim}}$) and trajectory accuracy (ATE, RPE-T, RPE-R). All methods use official code with identical inputs. $\uparrow$: Higher is better, $\downarrow$: Lower is better. Our method achieves the \textbf{best} or \underline{second-best} results.}
\vspace{-0.5em}
\label{tab:static}
\end{table}

\begin{table}[!t]
\centering
\scriptsize
\setlength{\tabcolsep}{0.5pt} %
\renewcommand{\arraystretch}{1}
\begin{tabular}{l cc ccc c} %
\toprule
& \multicolumn{2}{c}{\textbf{Generation Quality}} & \multicolumn{3}{c}{\textbf{Trajectory Accuracy}} & \multirow{2}{*}{\textbf{\makecell[c]{Training \\ -Free}}} \\
\cmidrule(lr){2-3}\cmidrule(lr){4-6}
& FVD $\downarrow$ & $\mathrm{CLIP\text{-}V}_{\mathrm{sim}} \uparrow$ 
& ATE $\downarrow$ & RPE-T $\downarrow$ & RPE-R $\downarrow$ & \\
\midrule
\makecell[l]{ViewExtrapolator \citep{ViewExtrapolator}} 
& 108.48 & 0.913
& 1.040 & 1.208 & 4.750
& \textcolor{green}{\ding{51}} \\ 
\makecell[l]{TrajectoryAttention \citep{TrajectoryAttention}} 
& 106.94 & 0.911
& 0.605 & 1.238 & \underline{3.560}
& \textcolor{red}{\ding{55}} \\ 
\makecell[l]{TrajectoryCrafter \citep{TrajCrafter}} 
& \underline{97.31} &\underline{0.923}
& \textbf{0.431} & \underline{1.078} & 8.950
& \textcolor{red}{\ding{55}} \\ 
\midrule
\makecell[l]{\textbf{WorldForge (Ours)}}
& \textbf{93.17} & \textbf{0.938}
& \underline{0.527} & \textbf{0.826} & \textbf{2.690}
& \textcolor{green}{\ding{51}} \\ 
\bottomrule
\end{tabular}
\caption{Quantitative comparison with existing methods on 4D dynamic scenes. We evaluate generation quality (FVD, $\mathrm{CLIP\text{-}V}_{\mathrm{sim}}$) and trajectory accuracy (ATE, RPE-T, RPE-R). All methods use official code with identical inputs.}
\vspace{-0.5em}
\label{tab:dynamic}
\end{table}

\begin{figure}[!t]
    \centering
\includegraphics[width=1\linewidth]{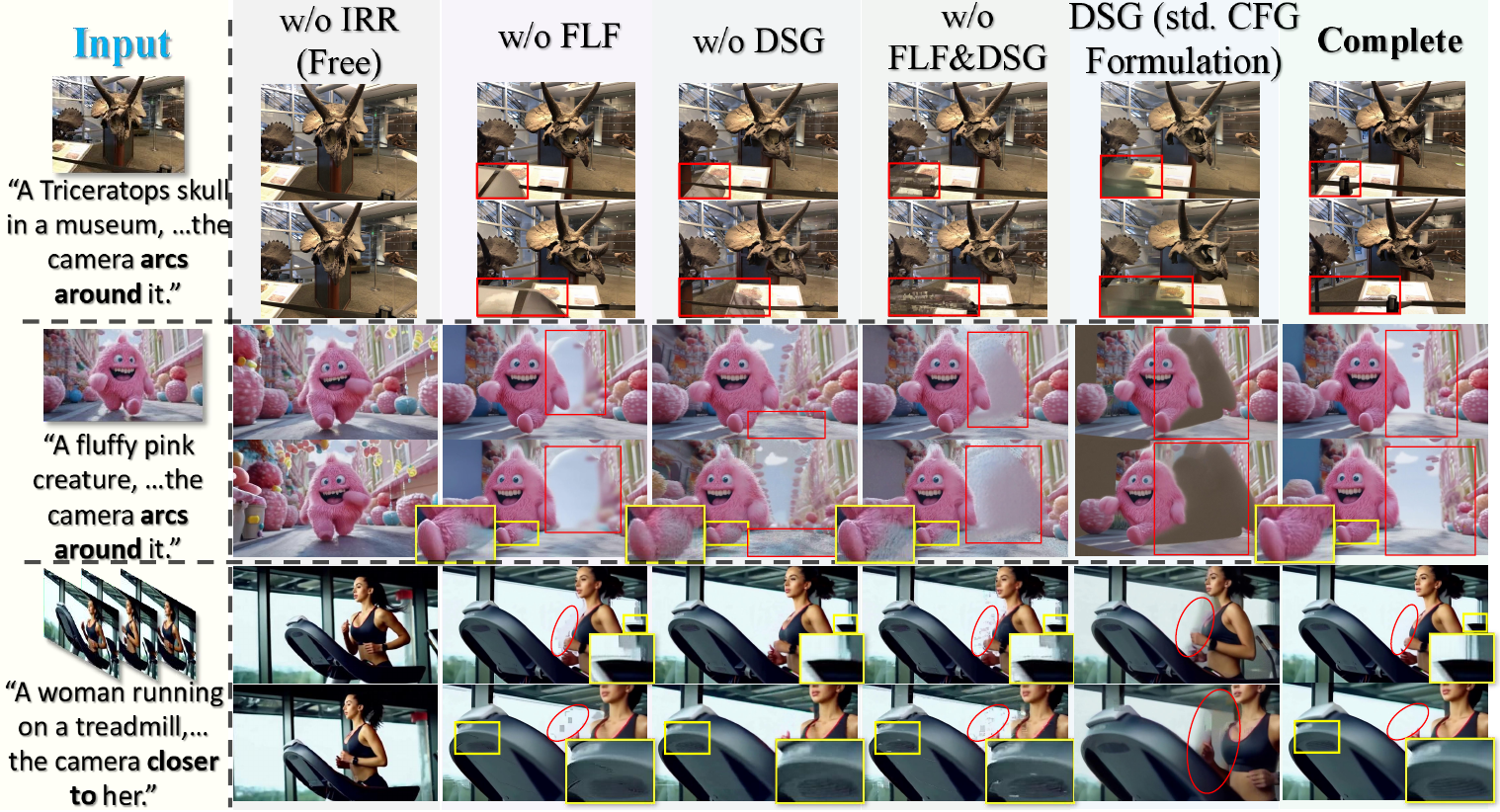}
    \caption{Ablation of the proposed components. IRR enables trajectory injection; without it, the model defaults to prompt-only free generation, and FLF/DSG cannot be applied. FLF decouples trajectory cues from noisy content; removing it introduces noise from warped frames. DSG guides sampling toward high-quality, trajectory-consistent results; without it, detail and plausibility drop. If the standard CFG formulation is applied in DSG, the large angular difference between the two velocity fields causes severe artifacts and errors. The full model achieves the best fidelity and control, demonstrating their complementary effects.}
    \label{fig:ablation}
\end{figure}

\subsection{Ablation Experiments}
\label{sec:4.4}
\begin{table}[!t]
\centering
\scriptsize
\setlength{\tabcolsep}{2pt} %
\renewcommand{\arraystretch}{0.75}
\begin{tabular}{l cc cc}
\toprule
& \multicolumn{2}{c}{\textbf{Static}} & \multicolumn{2}{c}{\textbf{Dynamic}} \\
\cmidrule(lr){2-3}\cmidrule(lr){4-5}
& FID $\downarrow$ & $\mathrm{CLIP}_{\mathrm{sim}} \uparrow$ & FVD $\downarrow$ & $\mathrm{CLIP\text{-}V}_{\mathrm{sim}} \uparrow$ \\
\midrule
\makecell[l]{w/o DSG} & 109.43 & 0.943 & 95.69 & 0.937 \\
\makecell[l]{w/o FLF} & 112.69 & 0.945 & 99.79 & 0.932 \\
\makecell[l]{w/o DSG\&FLF} & 113.12 & 0.943 & 103.17 & 0.931 \\
\makecell[l]{DSG (Using CFG Formulation)} & 120.91 & 0.936 & 109.1 & 0.919 \\
\midrule
\makecell[l]{\textbf{Complete Model (Ours)}} & \textbf{96.08} & \textbf{0.948} & \textbf{93.17} & \textbf{0.938} \\
\bottomrule
\end{tabular}
\caption{Quantitative ablation study of our core components. We report generation quality metrics for both static (FID, $\mathrm{CLIP}_{\mathrm{sim}}$) and dynamic (FVD, $\mathrm{CLIP\text{-}V}_{\mathrm{sim}}$) scenes. All components are shown to be essential for the best performance.}
\vspace{-0.5em}
\label{tab:ablation}
\end{table}
\textbf{Component Analysis.} As shown in Fig.~\ref{fig:ablation}, we remove IRR, FLF, and DSG in turn. Removing IRR disables trajectory guidance at inference time, resulting in failure to follow the target path. Without FLF, i.e., lacking motion/appearance separation, model priors become entangled, leading to unnatural outputs. Removing DSG introduces noise from warped trajectories into the generation process, causing artifacts and degrading visual quality. Using the standard CFG formulation (Eq.~\ref{eq:cfg}) in DSG also fails due to the large angular divergence, causing severe artifacts and errors. The complete model yields the best results, showing that all components are essential and work synergistically to enable robust and precise control.

\begin{figure}[!t]
    \centering
    \includegraphics[width=1\linewidth]{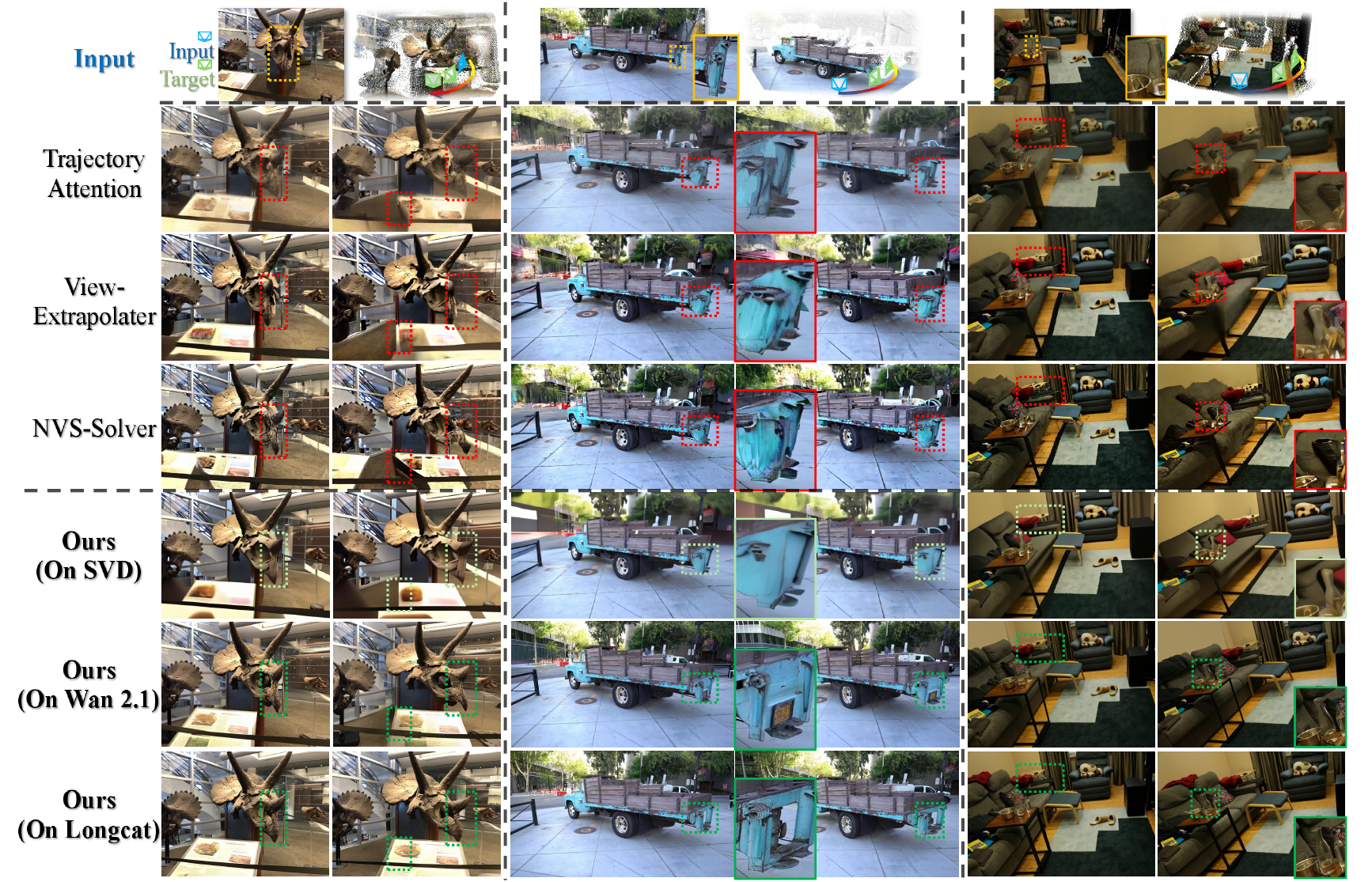}
    \caption{Ablation across different VDMs. To verify our method’s transferability, we port it to the U-Net–based SVD model \citep{SVD} and compare it against other SVD-based methods. Our guidance achieves excellent results on native SVD. Furthermore, we applied our method to the recent LongCat-Video~\citep{Longcat} model. Leveraging its rich world priors, our method again achieves SOTA results.}
    \label{fig:vdmablation}
    \vspace{-0.5em}
\end{figure}

\noindent\textbf{Video Model.} To test transferability, we integrated our method onto SVD \citep{SVD} and LongCat-Video~\citep{Longcat}. As shown in Fig. \ref{fig:vdmablation}, our method effectively adapts VDMs into trajectory-controllable scene generation models, its performance scaling positively with the base model's capabilities.

\noindent\textbf{Depth Model.} Relying on the VDM's strong world priors, our method effectively mitigates many warping-induced distortions. We experimented with different depth estimators \citep{VGGT, MegaSaM, UniDepth, DepthCrafter} and found our approach maintains robust performance, which allows for plug-and-play integration with various depth models. This integration demonstrates that our method will benefit from future advancements in depth estimation models.

More detailed comparative analysis of model performance, as well as efficiency analysis and metric definitions, are provided in the Supplementary.

\section{Conclusion}
\label{sec:conclusion}

We present \textbf{WorldForge}, a training-free framework for trajectory-controllable generation in static 3D and dynamic 4D scenes. Our method effectively balances visual quality, generalization, and precise control in video synthesis. At its heart is a unified, inference-time guidance strategy—comprising IRR, FLF, and DSG. By decoupling motion from appearance and correcting trajectory drift, our framework injects fine-grained control while preserving the rich world priors of the base model. Extensive experiments show state-of-the-art performance on both 3D and 4D generation tasks, offering a new path for exploring spatial intelligence in large-scale generative systems.

Although our framework generates high-quality static and dynamic scenes, it currently cannot meet real-time requirements due to its iterative guidance. Future work will focus on applying our method to more powerful generative models and distilling existing ones. This will aim to generate high-resolution, trajectory-controlled video sequences in just a few sampling steps.
\section*{Acknowledgement}

This work was supported by the National Natural Science Foundation of China (No. 6250070674) and the Zhejiang Leading Innovative and Entrepreneur Team Introduction Program (2024R01007).
{
    \small
    \bibliographystyle{ieeenat_fullname}
    \bibliography{main}
}

\clearpage
\setcounter{page}{1}
\setcounter{figure}{0}
\setcounter{table}{0}
\setcounter{equation}{0}
\setcounter{section}{0}
\maketitlesupplementary
\setcounter{tocdepth}{2} %
\tableofcontents         %
\vspace{0.5cm}             %
\addtocontents{toc}{\protect\setcounter{tocdepth}{2}}
\section{Proof of The Equivalence between Diffusion and Flow Models}
\label{Appendix_proof}

We consider Flow Matching \citep{Flow,flowstraight} as a special case of diffusion modeling \citep{understanding,DiffusionFlow}. In the following, we will first outline the formulation of diffusion models and then substitute the specific parameterization of Flow Matching to demonstrate their compatibility.

Given a random variable $\mathbf{x}_0$ drawn from an unknown data distribution $q_0(\mathbf{x}_0)$, a Diffusion Probabilistic Model (DPM) \citep{DDPM,PFODE,dpmsolver} defines a forward process that gradually transforms the data into a simple prior distribution, typically a Gaussian distribution. The conditional distribution of the noised variable $\mathbf{x}_t$ at time $t$ given the initial data $\mathbf{x}_0$ is defined as a Gaussian transition kernel \citep{vdm}:
\begin{equation}
    q_{t}(\mathbf{x}_t|\mathbf{x}_0)=\mathcal{N}(\mathbf{x}_t|\alpha_t\mathbf{x}_0,\sigma_t^2\mathbf{I}).
    \label{distribution}
\end{equation}
Equivalently, a sample $\mathbf{x}_t$ at any time $t \in [0, T]$ can be expressed through a reparameterization \citep{vdm,DiffusionFlow}:
\begin{equation}
    \mathbf{x}_t = \alpha_t\mathbf{x}_0 + \sigma_t \bm{\epsilon}, ~\bm{\epsilon} \sim \mathcal{N}(\mathbf{0},\mathbf{I}).
    \label{xt}
\end{equation}
Here, $\alpha_t$ and $\sigma_t$ are scalar functions of time, known as the noise schedule, that control the signal-to-noise ratio. Typically, $\alpha_t$ decreases over time while $\sigma_t$ increases, satisfying a condition such as $\alpha_t^2 + \sigma_t^2 = 1$ in Variance Preserving (VP) SDEs \citep{DDPM,PFODE}. \citet{vdm} proves that the following stochastic differential equation (SDE) has the same transition distribution in Eq.~(\ref{distribution}) for any $t \in [0,T]$:
\begin{equation}
\mathrm{d}\mathbf{x}_t=f(t)\mathbf{x}_t\mathrm{d}t+g(t)\mathrm{d}\mathbf{w}_t, ~\mathbf{x}_0\sim q_0(\mathbf{x}_0),
\end{equation}
where $\mathbf{w}_t$ is a standard Wiener process. The drift coefficient $f(t)$ and the diffusion coefficient $g(t)$ can be derived using schedule parameters $\alpha_t$ and $\sigma_t$ \citep{vdm}:
\begin{equation}
    f(t)=\frac{\mathrm{d}\log\alpha_t}{\mathrm{d}t},\quad g^2(t)=\frac{\mathrm{d}\sigma_t^2}{\mathrm{d}t}-2\frac{\mathrm{d}\log\alpha_t}{\mathrm{d}t}\sigma_t^2.
    \label{ft}
\end{equation}
The generative process of diffusion models involves reversing this forward process. This can be achieved via a corresponding reverse-time SDE \citep{PFODE}. For more efficient generation, one can utilize the associated probability flow ordinary differential equation (PF-ODE), which shares the same marginal distributions as at each time $t$ as that of the SDE \citep{PFODE}. This PF-ODE is given by:
\begin{equation}
    \frac{\mathrm{d}\mathbf{x}_t}{\mathrm{d}t}=f(t)\mathbf{x}_t - \frac{1}{2} g^2(t)\nabla_{\mathbf{x}_t}\log p_t(\mathbf{x}_t).
    \label{ode}
\end{equation}
By relating the score function $\nabla_{\mathbf{x}_t}\log p_t(\mathbf{x}_t)$ to the noise term via $\nabla_{\mathbf{x}_t}\log p_t(\mathbf{x}_t) \approx -\frac{\bm{\epsilon}_\theta(\mathbf{x}_t,t)}{\sigma_t}$, where $\bm{\epsilon}_\theta$ is a neural network trained to predict the noise, the ODE becomes \citep{EDM,UniPC}:
\begin{equation}
    \frac{\mathrm{d}\mathbf{x}_t}{\mathrm{d}t}=f(t)\mathbf{x}_t+\frac{g^2(t)}{2\sigma_t}\bm{\epsilon}_\theta(\mathbf{x}_t,t).
    \label{ode_simplified}
\end{equation}
Now, let us consider the forward process in Flow Matching \citep{Flow,flowstraight}. The path from a data point $\mathbf{x}_0$ to a noise sample $\bm{\epsilon}$ is defined by a simple linear interpolation:
\begin{equation}
    \mathbf{x}_t = (1-t)\mathbf{x}_0 + t \cdot \bm{\epsilon}, ~\bm{\epsilon} \sim \mathcal{N}(\mathbf{0},\mathbf{I}),
    \label{fm}
\end{equation}
where $t \in [0, 1]$. By comparing Eq.~(\ref{fm}) with the general form of the diffusion forward process in Eq.~(\ref{xt}), we can establish a direct correspondence by setting the diffusion schedule parameters as:
$$ \alpha_t = 1-t \quad \text{and} \quad \sigma_t = t. $$
Substituting this specific parameterization into the definitions for $f(t)$ and $g(t)$ in Eq.~(\ref{ft}), we derive the corresponding coefficients for this Flow Matching SDE:
\begin{equation}
    f_\mathrm{FM}(t)=\frac{\mathrm{d}\log(1-t)}{\mathrm{d}t} = \frac{-1}{1-t},
\end{equation}
\begin{equation}
    g^2_\mathrm{FM}(t)=\frac{\mathrm{d}(t^2)}{\mathrm{d}t}-2\frac{-1}{1-t}t^2 = \frac{2t}{1-t}.
\end{equation}
Next, we insert these specific coefficients $f_\mathrm{FM}(t)$ and $g^2_\mathrm{FM}(t)$ into the PF-ODE formulation from Eq.~(\ref{ode_simplified}). To analyze the underlying dynamics, we consider the ideal case where the score is perfectly known, which is equivalent to replacing the model prediction $\bm{\epsilon}_\theta(\mathbf{x}_t,t)$ with the ground-truth noise $\bm{\epsilon}$. This yields:
\begin{align}
    \frac{\mathrm{d}\mathbf{x}_t}{\mathrm{d}t} & =f_\mathrm{FM}(t)\mathbf{x}_t+\frac{g_\mathrm{FM}^2(t)}{2\sigma_t}\bm{\epsilon} \nonumber \\
    & = \frac{-1}{1-t}\mathbf{x}_t + \frac{2t}{2t \cdot (1-t)}\bm{\epsilon} \nonumber \\
    & = \frac{\bm{\epsilon}-\mathbf{x}_t}{1-t} \nonumber \\
    & = \frac{\bm{\epsilon}-[(1-t)\mathbf{x}_0 + t \cdot \bm{\epsilon}]}{1-t} \nonumber \\
    & = \frac{(1-t)\bm{\epsilon}-(1-t)\mathbf{x}_0}{1-t} \nonumber \\
    & = \bm{\epsilon} - \mathbf{x}_0.
    \label{diffusionflow}
\end{align}
This resultant vector field, $\frac{\mathrm{d}\mathbf{x}_t}{\mathrm{d}t} = \bm{\epsilon} - \mathbf{x}_0$, is precisely the time derivative of the Flow Matching path defined in Eq.~(\ref{fm}). This equivalence demonstrates that the process prescribed by Flow Matching is a specific instance of the diffusion models, corresponding to the linear noise schedule $\alpha_t = 1-t$ and $\sigma_t = t$. Therefore, Flow Matching can be formally viewed as a subset of the broader diffusion modeling framework \citep{understanding,DiffusionFlow}.

\section{Evaluation Metrics}
\label{sec:Sup B}

We employ seven complementary metrics to comprehensively evaluate video generation quality: FID and $\mathrm{CLIP}_{\mathrm{sim}}$ similarity for static scenes, FVD and $\mathrm{CLIP\text{-}V}_{\mathrm{sim}}$ for dynamic scenes, and ATE, RPE-T, and RPE-R for camera trajectory consistency. These metrics provide objective quantitative assessment across multiple dimensions including image realism, semantic consistency, temporal coherence, and camera motion fidelity.

\subsection{Static Scene Evaluation}

\textbf{Fréchet Inception Distance (FID).} 
FID~\citep{FID} measures image generation quality by comparing the distribution of real and generated images in the Inception-V3 feature space. We use an ImageNet-pretrained Inception-V3 \citep{Inception} model and extract 2048-dimensional features from the pool3 layer. The FID score is computed as:
\begin{equation}
\text{FID} = \|\mu_r - \mu_g\|^2 + \text{Tr}(\Sigma_r + \Sigma_g - 2(\Sigma_r\Sigma_g)^{1/2})
\end{equation}
where $\mu_r$ and $\mu_g$ are the mean vectors of real and generated image features, and $\Sigma_r$ and $\Sigma_g$ are the corresponding covariance matrices.

\textbf{CLIP Similarity.} 
CLIP similarity~\citep{CLIP} evaluates the semantic similarity between generated and real images using vision-language pre-trained representations. We employ the CLIP ViT-B/32 model trained on 400 million image-text pairs. The similarity score is calculated as:
\begin{equation}
\text{CLIP}_{\text{sim}} = \frac{1}{N} \sum_{i=1}^{N} \cos(f_{r,i}, f_{g,i})
\end{equation}
where $f_{r,i}$ and $f_{g,i}$ are the L2-normalized 512-dimensional CLIP features of the $i$-th real and generated image pair.

\subsection{Dynamic Scene Evaluation}

\textbf{Fréchet Video Distance (FVD).} 
FVD~\citep{FVD} measures distributional differences between real and generated video using pretrained spatio\mbox{-}temporal features. 
We use an I3D (Inflated 3D ConvNet) pretrained on Kinetics \citep{I3D} and extract 1024\mbox{-}D features from the global average pooling layer for each video clip. 
Following FID, we compute the Fréchet distance between the Gaussian fits of real and generated I3D features:
\begin{equation}
\text{FVD}=\|\mu_r-\mu_g\|_2^2+\mathrm{Tr}\!\left(\Sigma_r+\Sigma_g-2(\Sigma_r\Sigma_g)^{1/2}\right),
\end{equation}
with $\mu_r,\mu_g$ and $\Sigma_r,\Sigma_g$ estimated over clip\mbox{-}level I3D features.

\textbf{Video CLIP Similarity ($\mathrm{CLIP\text{-}V}_{\mathrm{sim}}$).} 
$\mathrm{CLIP\text{-}V}_{\mathrm{sim}}$ extends CLIP similarity to the temporal domain by computing frame-level semantic consistency between generated and real videos. The score is calculated as:
\begin{equation}
\mathrm{CLIP\text{-}V}_{\mathrm{sim}} = \frac{1}{M} \sum_{j=1}^{M} \left[ \frac{1}{T_j} \sum_{t=1}^{T_j} \cos(f_{r,j,t}, f_{g,j,t}) \right]
\end{equation}
where $M$ is the number of video pairs, $T_j$ is the frame count of the $j$-th video pair, and $f_{r,j,t}$, $f_{g,j,t}$ are the CLIP features of the $t$-th frame in the $j$-th video pair.

\subsection{Camera Trajectory Evaluation}

\textbf{Absolute Trajectory Error (ATE).} 
Before evaluation, we align the estimated trajectory to the reference by a global Sim3 transform (scale, rotation, translation). Let the aligned pose components be $\tilde{\mathbf{t}}_{\text{est},i}$ and $\tilde{\mathbf{R}}_{\text{est},i}$. 
ATE measures global consistency by the Euclidean distance between corresponding camera positions:
{\small
\begin{equation}
\begin{split}
\text{ATE}_i = \big\|\mathbf{t}_{\text{ref},i} - \tilde{\mathbf{t}}_{\text{est},i}\big\|_2,\\
\text{ATE} = \sqrt{\frac{1}{n} \sum_{i=1}^{n} \text{ATE}_i^2}.
\end{split}
\end{equation}
}
\textbf{Relative Pose Error — Translation (RPE-T).} 
RPE-T evaluates local translation accuracy between consecutive frames. 
Define relative motions via poses (index gap $\Delta\!=\!1$):
{\small
\begin{equation}
\Delta \mathbf{T}_{\text{ref},i}=\mathbf{T}_{\text{ref},i}^{-1}\mathbf{T}_{\text{ref},i+1}, 
\qquad 
\Delta \mathbf{T}_{\text{est},i}=\tilde{\mathbf{T}}_{\text{est},i}^{-1}\tilde{\mathbf{T}}_{\text{est},i+1}.
\end{equation}
}
Let $\Delta\mathbf{t}_{\text{ref},i}$ and $\Delta\mathbf{t}_{\text{est},i}$ be the translation parts of these relative transforms. The per-step error and RMSE are:
{\small
\begin{equation}
\begin{split}
\text{RPE-T}_i &= \big\|\Delta\mathbf{t}_{\text{ref},i} - \Delta\mathbf{t}_{\text{est},i}\big\|_2, \\
\text{RPE-T} &= \sqrt{\frac{1}{n-1} \sum_{i=1}^{n-1} \text{RPE-T}_i^2}.
\end{split}
\end{equation}
}
\textbf{Relative Pose Error — Rotation (RPE-R).} 
RPE-R assesses the accuracy of orientation changes between consecutive frames. 
Let the relative rotations be
{\small
\begin{equation}
\Delta\mathbf{R}_{\text{ref},i}=\mathbf{R}_{\text{ref},i}^{-1}\mathbf{R}_{\text{ref},i+1}, 
\qquad \!\!\!\!
\Delta\mathbf{R}_{\text{est},i}=\tilde{\mathbf{R}}_{\text{est},i}^{-1}\tilde{\mathbf{R}}_{\text{est},i+1}.
\end{equation}
}
The per-step angular error (degrees) and RMSE are:
{\small
\begin{equation}
\begin{split}
\text{RPE-R}_i = \arccos&\left(\frac{\mathrm{trace}\big(\Delta\mathbf{R}_{\text{ref},i}^\top \Delta\mathbf{R}_{\text{est},i}\big)-1}{2}\right)\cdot \frac{180}{\pi}, \\
\text{RPE-R} &= \sqrt{\frac{1}{n-1} \sum_{i=1}^{n-1} \text{RPE-R}_i^2}.
\end{split}
\end{equation}
}

\subsection{Evaluation Details}

\textbf{Preprocessing.} 
For FID, images are resized to $299\times299$ and fed to Inception\mbox{-}V3 with standard ImageNet normalization. 
For FVD and CLIP-based metrics, frames are resized to $224\times224$ with the respective model normalizations. 
To align with I3D input requirements for FVD, we uniformly downsample the generated videos to $16$ frames while strictly preserving the start and end frames to maintain boundary constraints. A similar temporal sampling strategy is applied for CLIP-based video metrics.
For camera trajectory evaluation, images are resized to $720\times480$ and uniformly sampled to $20$ frames.

\noindent\textbf{Evaluation Protocol.} 
To ensure robust feature covariance estimation for distributional metrics, we compute statistics over the aggregated evaluation set. 
Specifically, for FID on static scenes, we aggregate approximately $1,200$ GT images from $40+$ scenes as the reference distribution, comparing them against $\approx 2,200$ generated novel views (40$\sim$120 views per scene). 
For wild scenes without ground truth, we report the average FID over 5 representative scene groups using manually curated reference sets (20$\sim$40 images per scene) to ensure content consistency.
For FVD, we evaluate $\approx 700$ generated video clips from $50+$ scenes against an equal number of reference clips extracted from DAVIS and cinematic sequences.

All baseline methods are evaluated under this identical protocol. This standardized comparison ensures that the observed relative performance gaps reliably reflect the intrinsic differences in generation quality, verifying the superiority of our method across the evaluated metrics.

For trajectories, poses are recovered by SfM, the estimated trajectory is aligned to the reference by Sim3 to resolve scale, and metrics are computed using \texttt{evo} with alignment and scale correction enabled.

\section{Implementation Details}

This section provides a detailed breakdown of the Flow-Gated Latent Fusion (FLF) module,as introduced in Section~3.3 of the \emph{main} paper. The goal of FLF is to identify and selectively update latent channels that are highly relevant to motion, thereby preserving visual details encoded in appearance-focused channels. To achieve this, at each denoising step $i$, FLF computes a motion similarity score $S^{(t,c)}$ for each latent channel $c$. Below, we detail how this score is calculated.

\subsection{Details for FLF Estimation and Motion Scoring}
\paragraph{Optical Flow Estimation}
At each denoising step $i$, we compute optical flow maps for each channel $c$ of both the predicted latent $\hat{\mathbf{x}}_0^{(t)}$ and the target trajectory latent $\mathbf{x}_{\text{traj}}$. The computation is performed frame-by-frame; that is, for each latent tensor, we calculate the dense optical flow between consecutive temporal frames using the Farnebäck algorithm~\citep{OpticaFlow}. This process yields a predicted flow map, $\mathcal{F}^{(t,c)}_{\text{pred}}$, and a ground-truth (GT) flow map, $\mathcal{F}^{(t,c)}_{\text{gt}}$. At each pixel, the flow is a 2D vector $(u_*, v_*)$ representing horizontal and vertical displacement. All subsequent metric calculations are performed over the set of valid (i.e., non-occluded) pixels, defined as $\Omega^{(t,c)} = \{(x,y,\tau) \mid \mathbf{M}^{(c)}(x,y,\tau)=1\}$, where $(x,y)$ are pixel coordinates and $\tau$ is the frame index. Since optical flow is computed between adjacent frames, for a latent tensor with $T_l$ total frames, the index $\tau$ ranges from $1$ to $T_{l}-1$.

\paragraph{Metric Calculation}
The motion score $S^{(t,c)}$ is derived from three standard optical flow metrics that quantify the error between the predicted flow $\mathcal{F}^{(t,c)}_{\text{pred}}$ and the ground-truth flow $\mathcal{F}^{(t,c)}_{\text{gt}}$ at each step $i$.

\begin{itemize}
    \item \textbf{Masked End-point Error (M-EPE)} measures the average Euclidean distance between the predicted and GT flow vectors over all valid pixels. Let $err(x,y,\tau)$ be the Euclidean error at a specific pixel:
    {\small
    \begin{equation}
        err(x,y,\tau) = \left\| \mathcal{F}^{(t,c)}_{\text{pred}}(x,y,\tau) - \mathcal{F}^{(t,c)}_{\text{gt}}(x,y,\tau) \right\|_2.
    \end{equation}}
    The M-EPE is then calculated as:
    {\small\begin{equation}
        \text{M-EPE}^{(t,c)} = \frac{1}{|\Omega^{(t,c)}|} \sum_{(x,y,\tau) \in \Omega^{(t,c)}} err(x,y,\tau).
    \end{equation}}

    \item \textbf{Masked Angular Error (M-AE)} calculates the average angular difference. We first define the cosine similarity $sim(x,y,\tau)$ between the flow vectors:
    {\small\begin{equation}
        sim(x,y,\tau) = \frac{ \mathcal{F}^{(t,c)}_{\text{pred}}(x, y,\tau) \cdot  \mathcal{F}^{(t,c)}_{\text{gt}}(x, y,\tau)}{\| \mathcal{F}^{(t,c)}_{\text{pred}}(x, y,\tau)\| \cdot \| \mathcal{F}^{(t,c)}_{\text{gt}}(x, y,\tau)\|}.
    \end{equation}}
    The M-AE is derived by averaging the arccosine of this similarity:
    {\small\begin{equation}
        \text{M-AE}^{(t,c)} = \frac{1}{|\Omega^{(t,c)}|} \!\!\sum_{(x,y,\tau) \in \Omega^{(t,c)}} \!\!\!\!\arccos\left( sim(x,y,\tau) \right).
    \end{equation}}

    \item \textbf{Outlier Percentage (Fl-all)} is the percentage of pixels in $\Omega^{(t,c)}$ where the flow estimation is considered erroneous. Following standard benchmarks, a pixel is flagged as an outlier if its M-EPE exceeds 3 pixels or if its relative error, $\| \mathcal{F}^{(t,c)}_{\text{pred}} - \mathcal{F}^{(t,c)}_{\text{gt}} \|_2 / \| \mathcal{F}^{(t,c)}_{\text{gt}} \|_2$, is greater than 5\%. We denote this outlier percentage as $\text{F}^{(t,c)}$.
\end{itemize}

\paragraph{Normalization and Weighting}
The three metrics exist on different scales, so we first normalize each to the range $[0, 1]$ before combining them. This corresponds to the ${\text{Norm}_k}^{(t,c)}$ terms used in the main text:
\begin{equation}
\begin{aligned}
{\text{Norm}_\text{E}}^{(t,c)} &= \min(\text{M-EPE}^{(t,c)}/n_{\text{E}}, 1), \\
{\text{Norm}_\text{A}}^{(t,c)} &= \min(\text{M-AE}^{(t,c)}/n_{\text{A}}, 1), \\
{\text{Norm}_\text{F}}^{(t,c)} &= \min(\text{F}^{(t,c)}/n_{\text{F}}, 1),
\end{aligned}
\end{equation}
where $n_{\text{E}}, n_{\text{A}},$ and $n_{\text{F}}$ are normalization constants chosen to reflect typical value ranges for each metric. The final motion score $S^{(t,c)}$ is a weighted sum of the inverted normalized errors, as defined in Eq.~(\ref{eq:score1}) (corresponding to Eq.~(6) in the \emph{main text}):
\begin{equation}
S^{(t,c)}=\sum_{k\in\{\text{E, A, F}\}} \gamma_k\!\left(1 - {\text{Norm}_k}^{(t,c)}\right),
\label{eq:score1}
\end{equation}
where the weights $\gamma_k$ (where $k\in\{\text{E},\text{A},\text{F}\}$ and $\sum_{k}\gamma_k=1$) and the normalization constants are set based on common practices in optical flow evaluation to balance each metric’s contribution. In our experiments, we set $n_{\text{E}}=10$, $n_{\text{A}}=30$, and $n_{\text{F}}=0.5$. The weights in Eq.~(\ref{eq:score1}) are set to $(\gamma_{\text{E}},\gamma_{\text{A}},\gamma_{\text{F}})=(0.4,0.4,0.2)$.

\begin{figure*}[htbp]
    \centering
    \includegraphics[width=0.95\linewidth]{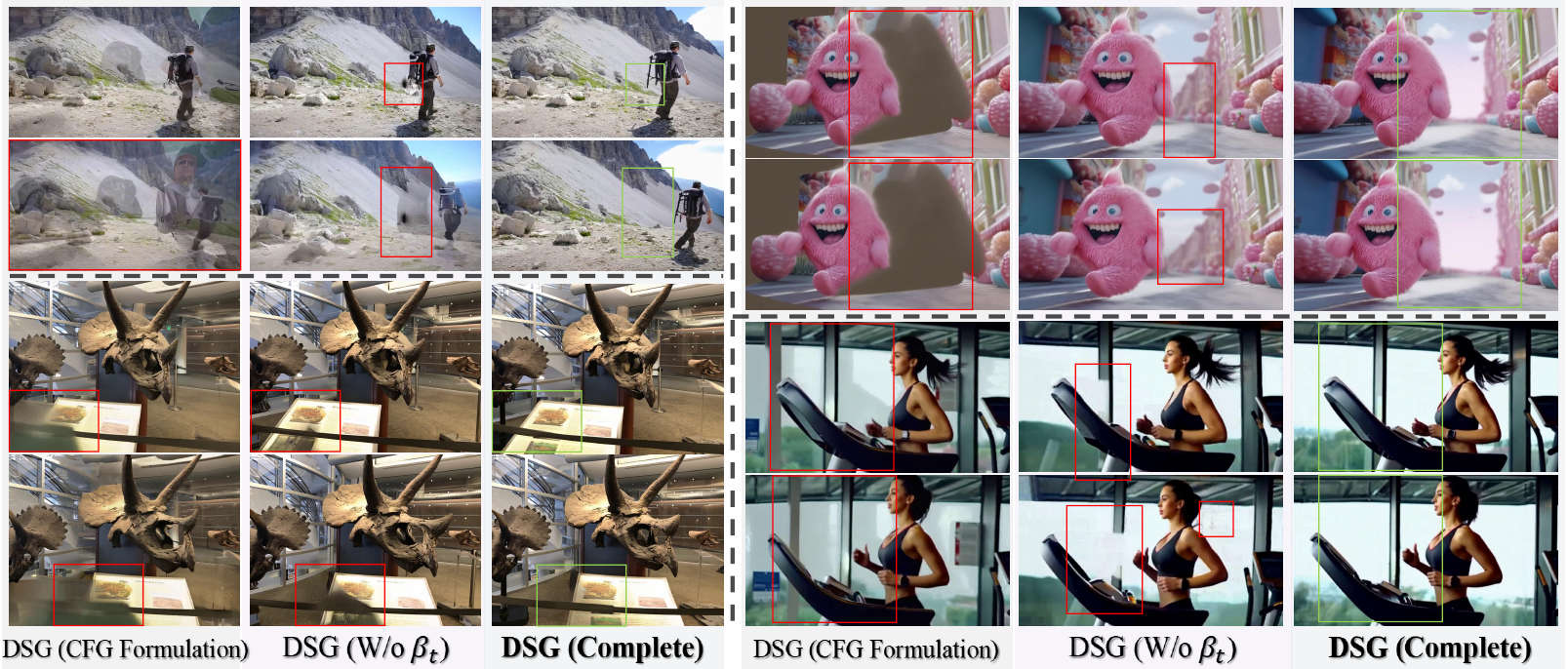}
\caption{Qualitative ablation study of the DSG method. Substituting DSG with a standard CFG formulation fails to handle the large angular disparity between the two velocity fields, resulting in significant visual artifacts and errors. Removing the adaptive weighting factor $\beta_t$ (denoted as DSG w/o $\beta_t$) compromises guidance stability and introduces inconsistencies. In contrast, our full DSG framework stably generates high-fidelity and structurally consistent results.}
    \label{fig:dsg_vs_cfg}
\end{figure*}
\subsection{Hyperparameter Settings}
\label{sec:hyperparams}

To facilitate reproducibility, we provide the default hyperparameter settings used in our experiments, as listed in Table~\ref{tab:param}. These values are based on the implementation with the Wan 2.1 backbone. In practical applications, users may fine-tune these coefficients according to specific scene requirements to achieve optimal results.

\begin{table}[htbp]
\centering
\small
\setlength{\tabcolsep}{1pt}
\renewcommand{\arraystretch}{1.2}
\caption{Default coefficient settings used in our experiments (taking Wan 2.1 implementation as an example). While these values serve as a robust baseline, users can fine-tune them for specific scenes to maximize generation quality.}
\begin{tabular}{lccc}
\toprule
 & \textbf{Reference} & \textbf{Value} & \textbf{Description} \\
\midrule
$\gamma_k$ & \makecell{Eq. (6) \\in \emph{main} paper} & $[0.4, 0.4, 0.2]$ & Weights for optical flow metrics \\
$\lambda^{(t)}$ & \makecell{Eq. (6) \\in \emph{main} paper} & $0.65$ & \makecell{Threshold for FLF \\channel filtering} \\
$\rho$ &\makecell{Eq. (8) \\in \emph{main} paper} & $4.0$ & DSG guidance scale \\
\bottomrule
\end{tabular}
\label{tab:param}
\end{table}

Beyond the coefficients listed above, we specify several key operational parameters. Taking the Wan 2.1 model (which uses 50 sampling steps) as an example, the Intra-Step Recursive Refinement (IRR) is applied by default during the first 20 sampling steps. For optical flow estimation within the FLF module, we utilize the classic Farneback algorithm~\citep{OpticaFlow} with its default settings (`pyr\_scale`=0.5, `levels`=3, `winsize`=15).

Critically, regarding the channel filtering in FLF, in addition to the threshold $\lambda^{(t)}$, we implement a progressively relaxed channel selection strategy to balance structural guidance and detail preservation:
\textbf{Phase 1 (Steps 0--5):} Channel filtering is disabled. All channels are injected with guidance information. This is because the early-stage latent states are too noisy for reliable optical flow calculation, necessitating full guidance injection to establish initial structure.
 \textbf{Phase 2 (Steps 6--10):} We enforce a strict limit where a maximum of 2 channels are allowed to retain the original prediction (i.e., at least 14 channels are replaced with guided features). This ensures sufficient structural information is injected during the formative stages of generation.
\textbf{Phase 3 (Steps $\ge$ 11):} The constraint is relaxed to allow a maximum of 6 channels to retain the original prediction. This looser constraint prevents excessive guidance from compromising fine texture details in the later stages.

\begin{figure*}[!t]
    \centering
    \includegraphics[width=0.9\linewidth]{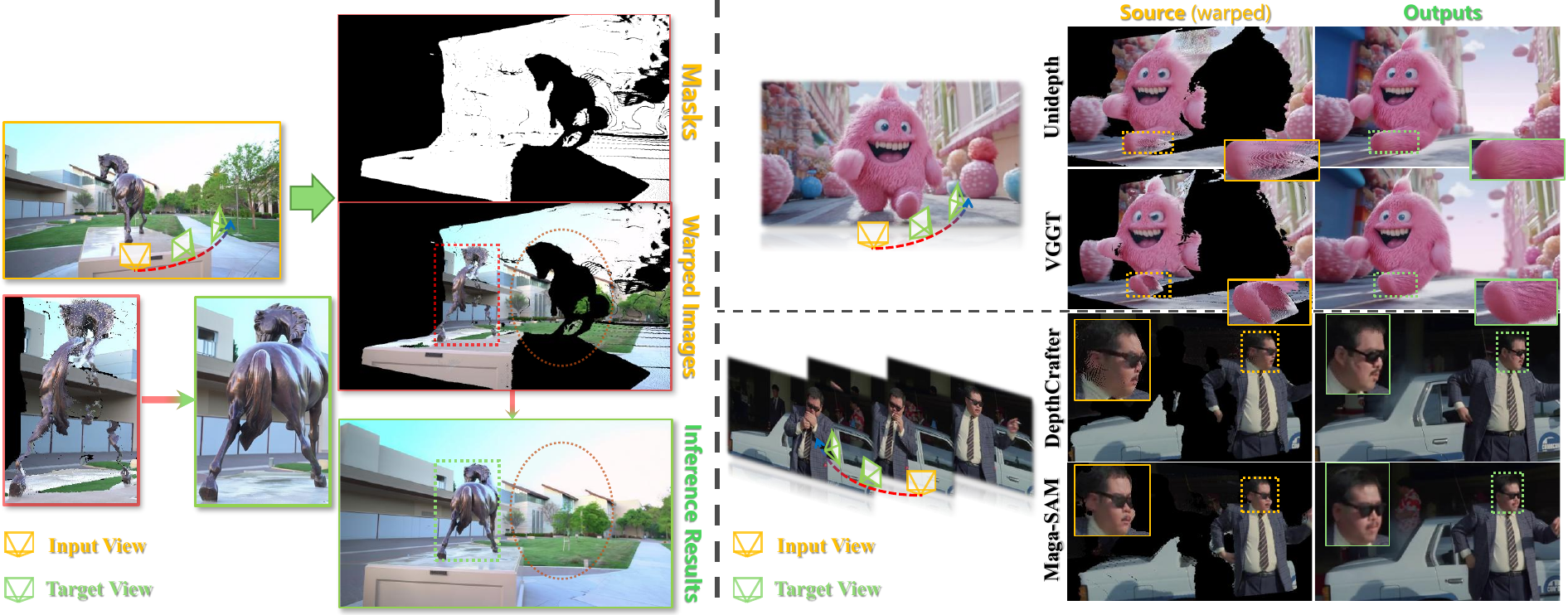}
    \caption{Depth-models ablation. Our method leverages the inherent world knowledge of VDMs to correct errors and fill missing regions even under challenging inputs (left). This strong self-correction ability ensures broad compatibility with different depth estimators (right). Despite variations or noise in depth-based warping, it reliably compensates through learned priors and produces realistic, high-quality results.}
    \label{fig:depthablation}
\end{figure*}

\begin{figure*}[!t]
    \centering
    \includegraphics[width=0.95\linewidth]{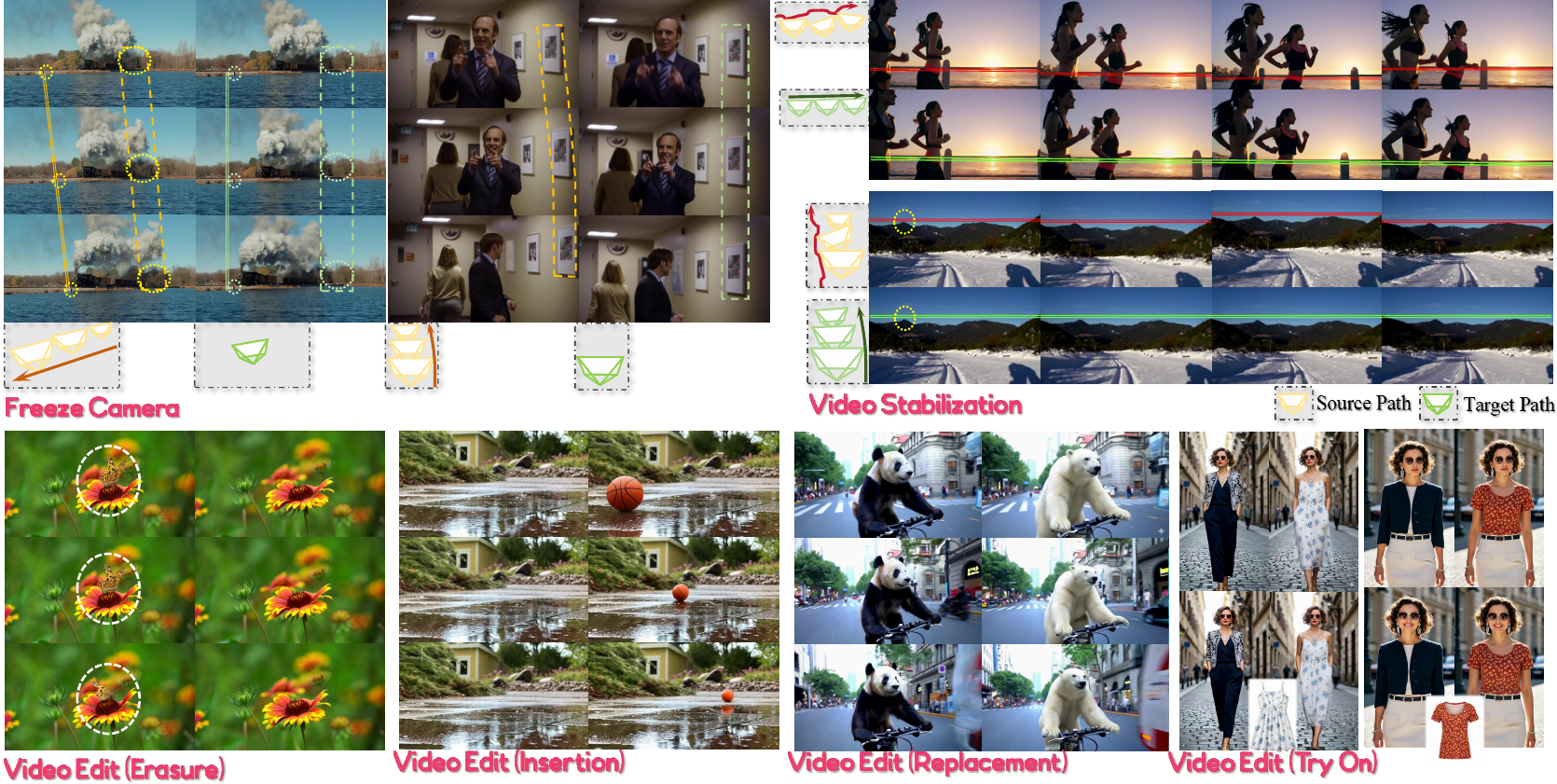}
    \caption{Other video effects enabled by our method. Beyond video re-cam, our flexible depth-based warping also supports various video editing operations, such as freezing the camera, stabilizing video, and editing video content. These extensions further broaden the practical scope of our approach.}
    \label{fig:edit}
\end{figure*}
\section{More Experimental Results}
\label{Appendix_Study}
This section provides additional experiments and results that complement the findings presented in the main paper. We include a detailed efficiency analysis, further ablation studies, and more qualitative examples to fully demonstrate the capabilities and robustness of our framework.

\subsection{Efficiency and Runtime Analysis}
\label{Appendix_time}
Our framework is training-free and operates entirely at inference time. Table~\ref{tab:time} provides a detailed comparison of inference efficiency against several state-of-the-art methods on a single NVIDIA A100 GPU.
\begin{table}[htbp]
\centering
\scriptsize
\setlength{\tabcolsep}{0.1pt}
\renewcommand{\arraystretch}{0.8}
\caption{Efficiency comparison. We measure inference throughput on a single NVIDIA A100 across methods built on SVD \citep{SVD}, Wan 2.1 \citep{Wan21}, CogVideoX \citep{CogVideoX}, LongCat \citep{Longcat}, and custom backbones. Our approach achieves competitive generation speed compared to prior approaches while avoiding any training overhead. \textbf{\emph{SR}} denotes LongCat's built-in super-resolution model, and \textbf{\emph{D}} represents its distilled version. The \textbf{best} and \underline{second-best} results are highlighted in bold and underlined, respectively.}
\begin{tabular}{lcccc}
\toprule
& \textbf{Resolution} & \textbf{\makecell{Inference \\Speed\\(frames/min)}} & \textbf{\makecell{Base Video\\ Model}} & \textbf{\makecell{Training\\\mbox{-}Free}} \\
\midrule
\makecell[l]{See3D \citep{See3D}}             & $576\times1024$   & 14.71  & Custom          & \textcolor{red}{\xmark} \\
\makecell[l]{ViewCrafter \citep{ViewCrafter}}            & $576\times1024$   & 13.89  & Custom          & \textcolor{red}{\xmark} \\
\makecell[l]{ViewExtrapolator \citep{ViewExtrapolator}}    & $576\times1024$   & 15.63  & SVD              & \textcolor{green}{\cmark} \\
\makecell[l]{TrajectoryAttention \citep{TrajectoryAttention}} & $576\times1024$ & 4.55   & SVD              & \textcolor{red}{\xmark} \\
\makecell[l]{TrajectoryCrafter \citep{TrajCrafter}}       & $384\times672$    & 14.71  & CogVideoX        & \textcolor{red}{\xmark} \\
\makecell[l]{NVS\mbox{-}Solver \citep{NVSSolver}}         & $576\times1024$   & 2.69   & SVD              & \textcolor{green}{\cmark} \\
\makecell[l]{ReCamMaster \citep{ReCamMaster}}            & $480\times832$    & 5.55   & Wan~2.1 T2V      & \textcolor{red}{\xmark} \\
\midrule
\makecell[l]{WorldForge, on Wan 2.1 \citep{Wan21}}                  & $720\times1280$   & 1.45   & Wan~2.1 I2V      & \textcolor{green}{\cmark} \\
\makecell[l]{WorldForge, on Wan 2.1 \citep{Wan21}}                  & $480\times832$    & 3.68   & Wan~2.1 I2V      & \textcolor{green}{\cmark} \\
\makecell[l]{WorldForge, on SVD \citep{SVD}}                  & $576\times1024$   & \textbf{19.23}  & SVD              & \textcolor{green}{\cmark} \\
\makecell[l]{WorldForge, on LongCat \citep{Longcat}}                   & $480\times832$    & 3.85   & Longcat          & \textcolor{green}{\cmark} \\
\makecell[l]{WorldForge, on LongCat \citep{Longcat}}              &\ $720\times1280$   & 3.33   & \makecell{Longcat (\textbf{\emph{SR})}} & \textcolor{green}{\cmark} \\
\makecell[l]{WorldForge, on LongCat \citep{Longcat}}         & $480\times832$    & \underline{16.67}  & \makecell{Longcat (\textbf{\emph{D})}}     & \textcolor{green}{\cmark} \\
\makecell[l]{WorldForge, on LongCat \citep{Longcat}}      & $720\times1280$   & 10.20  & \makecell{Longcat (\textbf{\emph{D}}+\textbf{\emph{SR})}}       & \textcolor{green}{\cmark} \\
\bottomrule
\end{tabular}
\label{tab:time}
\end{table}

Our method incurs zero training cost, offering a significant advantage over resource-intensive fine-tuning approaches. The primary computational overhead stems from the IRR module, which effectively adds an extra sampling process, taking approximately the same time as a single standard sampling step. In contrast, the DSG module involves only a simple matrix operation per step, incurring negligible temporal cost. Similarly, the FLF module's channel-wise optical flow estimation presents a substantially lower computational burden than the backbone model inference. A detailed breakdown of these component costs is reported in Table~\ref{tab:cost}. Note that while absolute times may vary across hardware configurations, the relative temporal relationships remain reliable. Despite these additions, our framework maintains inference speeds comparable to, and often faster than, existing methods, as evidenced in Table~\ref{tab:time}. This demonstrates that our framework achieves robust controllability without prohibitive computational costs, offering an efficient alternative to training-intensive pipelines.

\begin{table}[!t]
\centering
\scriptsize %
\setlength{\tabcolsep}{12pt} %
\renewcommand{\arraystretch}{1.2} %
\caption{Computational cost breakdown of a single generation step. We report the runtime of each component on an NVIDIA A100, taking the generation of a 49-frame video at $832\times480$ resolution as an example. By default, we apply our guidance during the first 20 sampling steps. The primary overhead comes from the IRR module, while the DSG module incurs negligible cost.}
\begin{tabular}{lc}
\toprule
\textbf{Component} & \textbf{Time (s)} \\
\midrule
VAE Encoding \& Decoding & 4.5 \\
Backbone Inference (Transformer) & 8.4 \\
\midrule
FLF Module (Ours) & 1.2 \\
DSG Module (Ours) & $\approx 0$ \\
IRR Module (Ours) & 14.1 \\
\midrule
Total Runtime & \textbf{29.0} \\
\bottomrule
\end{tabular}
\label{tab:cost}
\end{table}

\subsection{Ablation of DSG and Naive CFG}
\label{Appendix_DSG}
Through extensive experiments, we observe that the difference between our trajectory-guided velocity $\mathbf{v}_{t}^{\text{traj}}$ and the unguided velocity $\mathbf{v}_{t}^{\text{ori}}$ is far greater than that between the conditional $\mathbf{v}_{\text{con}}$ and unconditional $\mathbf{v}_{\text{uncon}}$ estimates in standard CFG~\citep{CFG}. Specifically, empirical analysis reveals that the angular difference in our setting typically ranges from $50^\circ$ to $70^\circ$, significantly exceeding the $<5^\circ$ divergence found in typical CFG scenarios. To mitigate the adverse effects caused by this large angular discrepancy, we propose Dual-Path Self-Corrective Guidance (DSG). A direct visual comparison is presented in Fig.~\ref{fig:dsg_vs_cfg}, where we compare our full framework against a standard CFG implementation and an ablated version without the adaptive weight $\beta_t$. The results demonstrate that replacing DSG with a naive CFG formulation leads to severe visual artifacts and structural distortions, while removing the adaptive weighting factor reduces guidance stability. In contrast, our full DSG framework successfully maintains structural integrity and high perceptual quality while closely adhering to the intended camera path.

\subsection{Ablation on Video Diffusion Models}
\label{Appendix_VDM}

To evaluate the transferability of our proposed guidance mechanism and its performance across models of varying parameter scales, we conducted ablation studies by porting our entire framework to the U-Net-based Stable Video Diffusion (SVD)~\citep{SVD}, which possesses fewer parameters, and the recently released LongCat model~\citep{Longcat}. We performed minor hyperparameter fine-tuning to adapt our method to their respective architectures and sampling strategies. Subsequently, we conducted a fair comparison using identical inputs. Quantitative results are presented in Table~\ref{tab:vdm}. It is worth noting that due to SVD's constraints in parameter count and architecture, it encapsulates fewer inherent world priors, which prevents it from fully exploiting the potential of our guidance algorithm. Comprehensive visualization results demonstrating the capabilities across these models are shown in Fig.~\ref{fig:more_3d_1} through Fig.~\ref{fig:more_4d_3}.
\begin{table}[htbp]
\centering
\footnotesize
\setlength{\tabcolsep}{3pt} %
\renewcommand{\arraystretch}{1}
\caption{Quantitative comparison across different backbones. Using single-view 3D scene generation as a benchmark, we evaluate our method on SVD~\citep{SVD}, Wan 2.1~\citep{Wan21}, and LongCat~\citep{Longcat}. The results demonstrate the scalability of our approach and its ability to generalize across different VDM architectures. Furthermore, the performance gains on advanced backbones indicate that our method effectively leverages the capabilities of the underlying model, promising improved generation quality as base models continue to evolve.}
\begin{tabular}{l c c c c}
\toprule
 & CLIP $\uparrow$ & ATE $\downarrow$ & RPE-T $\downarrow$ & RPE-R $\downarrow$ \\
\midrule
\makecell[l]{WorldForge (on SVD~\citep{SVD})}      & 0.910 & 0.265 & 0.316 & 0.444 \\
\makecell[l]{WorldForge (on Wan 2.1~\citep{Wan21})}  & 0.948 & \textbf{0.077} & 0.086 & \textbf{0.221} \\
\makecell[l]{WorldForge (on LongCat~\citep{Longcat})}  & \textbf{0.949} & 0.095 & \textbf{0.076} & 0.230 \\
\bottomrule
\end{tabular}
\label{tab:vdm}
\end{table}

\subsection{Ablation on Depth Estimation Models}
\label{Appendix_Depth}

Our framework operates on a warp-and-repaint strategy. To assess the robustness and flexibility of our approach regarding depth estimation, we evaluated its performance using several state-of-the-art depth estimators: VGGT~\citep{VGGT}, UniDepth~\citep{UniDepth}, Mega-SaM~\citep{MegaSaM}, and DepthCrafter~\citep{DepthCrafter}. As illustrated in Fig.~\ref{fig:depthablation}, our method demonstrates broad compatibility, maintaining consistently high performance across all tested models. Even when depth-based warping yields challenging inputs characterized by noise, errors, or significant disocclusion regions, our framework effectively compensates for these imperfections. This resilience stems from the strong generative world priors inherent in the underlying VDM, which our guidance modules leverage to correct artifacts and plausibly inpaint missing areas during the repainting stage. This self-correction capability confirms that our framework functions in a plug-and-play manner with various depth estimation techniques and naturally scales with improvements in depth estimation performance.

\subsection{Applications in Video Editing}
\label{Appendix_edit}
\begin{figure*}[!t]
    \centering
    \includegraphics[width=0.9\linewidth]{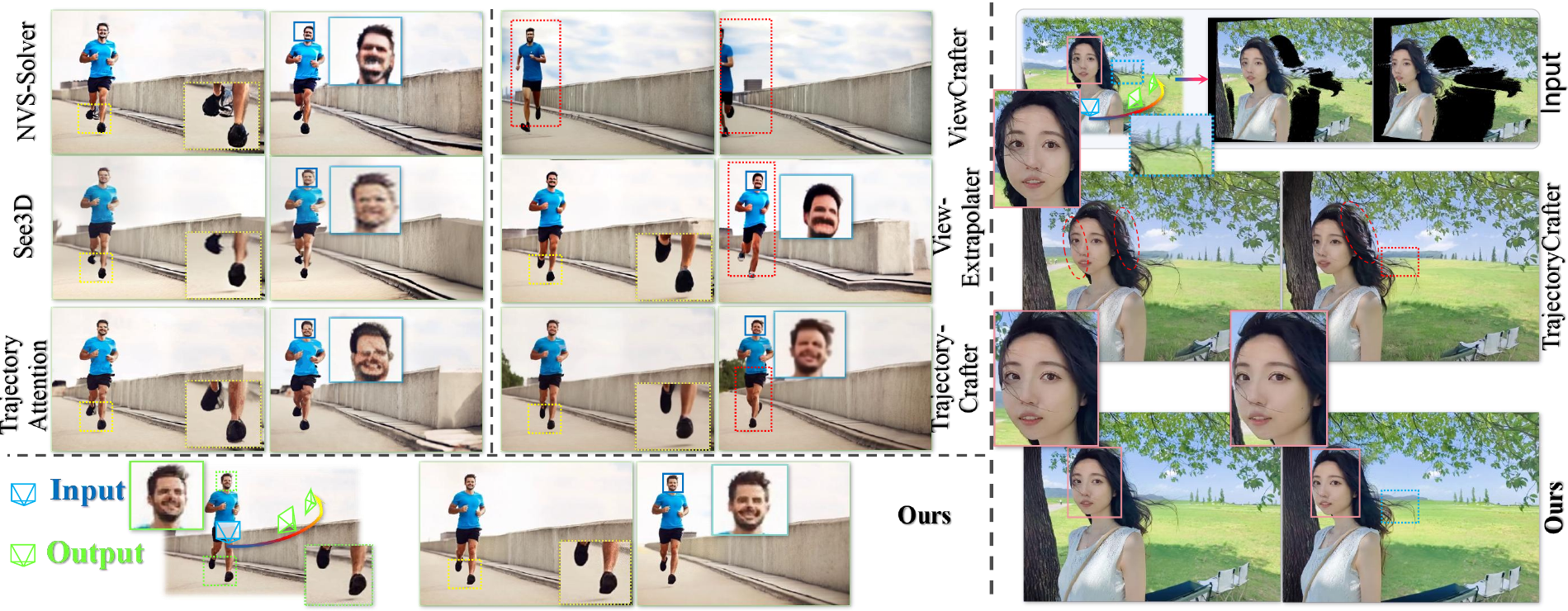}
    \caption{Static 3D generation on human-centric scenes. Existing methods struggle, particularly with motion-prone shots (left) and portrait close-ups (right). On the left, baselines introduce artifacts and unintended motion. On the right, most fail to produce plausible results; TrajectoryCrafter \citep{TrajCrafter} recovers coarse structure but lacks detail and visual appeal. In contrast, our method maintains scene stationarity under trajectory guidance and produces natural, faithful renderings, achieving both precise control and high perceptual quality.}
    \label{fig:man}
\end{figure*}
Beyond trajectory-controlled generation, our framework's flexibility makes it a powerful tool for various video post-production and editing tasks. This includes effects like video stabilization, camera freezing, and dynamic viewpoint switching.

Furthermore, by incorporating a flexible masking strategy, our framework can perform diverse content edits such as object removal, addition, subject replacement, and virtual try-on seamlessly. The general process for these edits involves first segmenting the target region in each frame using a tool like SAM~\citep{SAM}. The desired edit is then applied to the first frame (e.g., using Gemini~\citep{Gemini}). Finally, this edited frame and the corresponding masks are processed by our pipeline to render a temporally consistent result. For adding new objects where none exist in the source video, a simple bounding box can be provided to guide the placement. Fig.~\ref{fig:edit} shows several qualitative examples of these video editing effects.

\subsection{Generation on Challenging Scenes}
\label{Appendix_more}
\begin{figure*}[!t]
    \centering
    \includegraphics[width=0.9\linewidth]{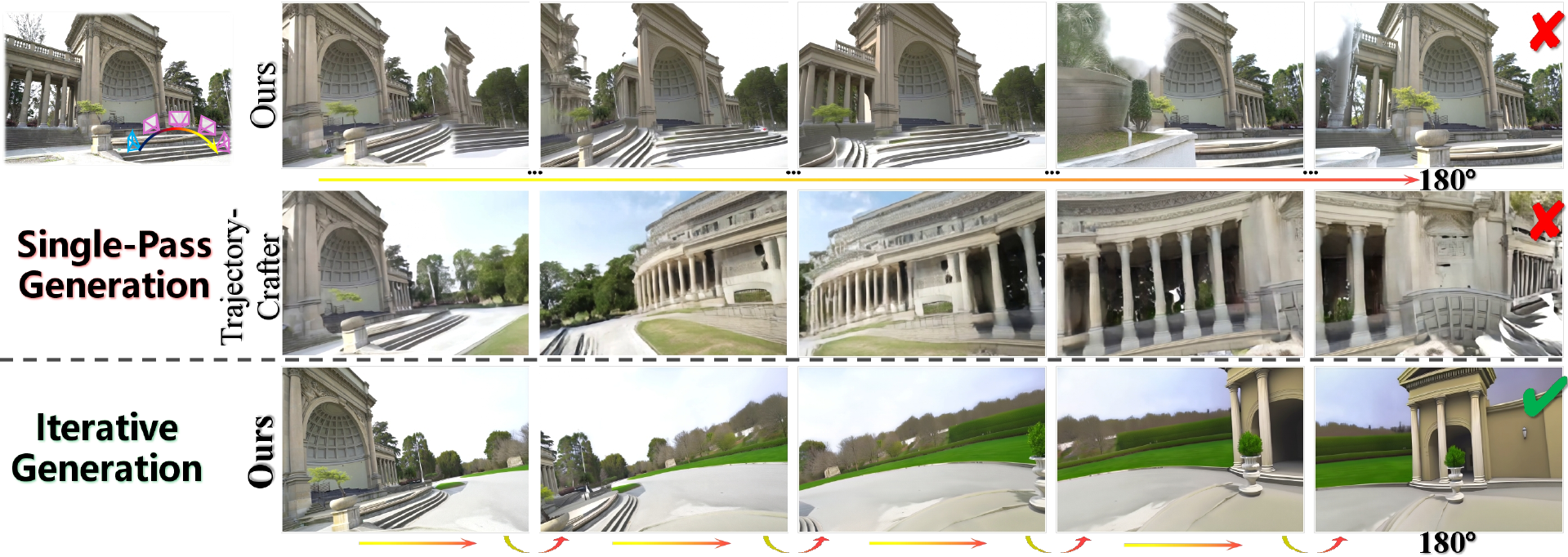}
    \caption{Large camera movements (e.g., $180^\circ$). Single-pass generation of large angles often suffers from poor quality. Our method effectively resolves this problem via iterative generation.}
    \label{fig_extreme_motion}
\end{figure*}

\begin{figure*}[!t]
    \centering
    \includegraphics[width=0.9\linewidth]{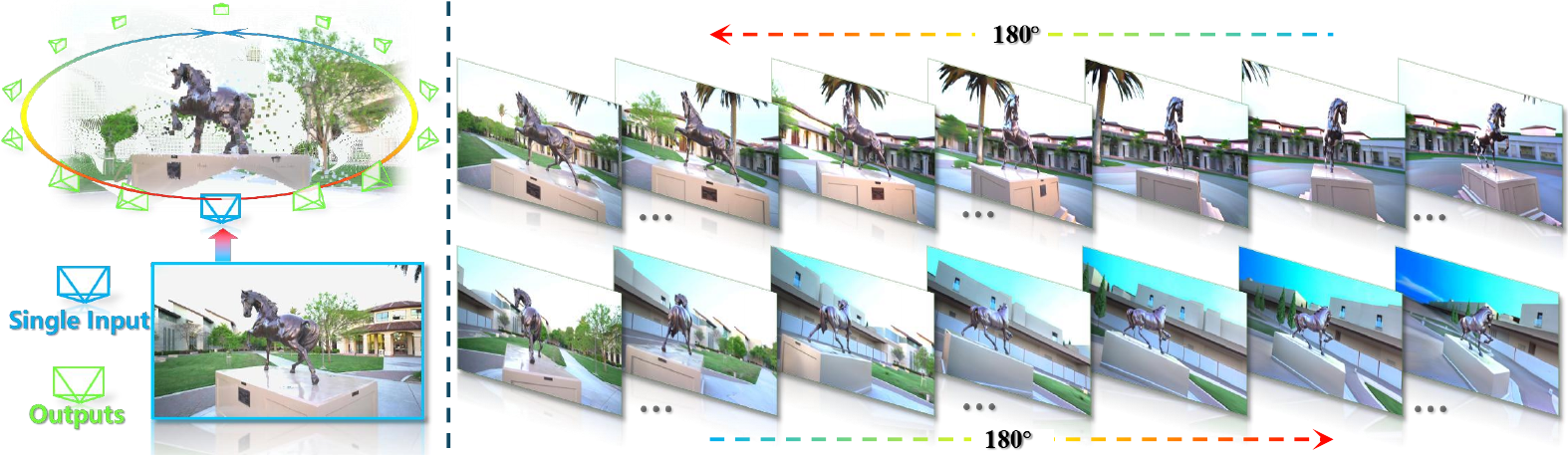}
    \caption{360$^\circ$ orbit views from a single real-world outdoor image. With precise trajectory control and realistic rendering, our method overcomes the viewpoint limitation of single-image generation and produces ultra-wide views of complex real scenes. Unlike panorama-based approaches, it directly supports object-centric trajectories and achieves higher visual quality.}
    \label{fig:360}
\end{figure*}
Our approach demonstrates robust performance in difficult cases where other methods may falter. We highlight two such scenarios: human-centric scenes and single-image 360° view generation.
\begin{figure*}[htbp]
    \centering
    \includegraphics[width=0.9\linewidth]{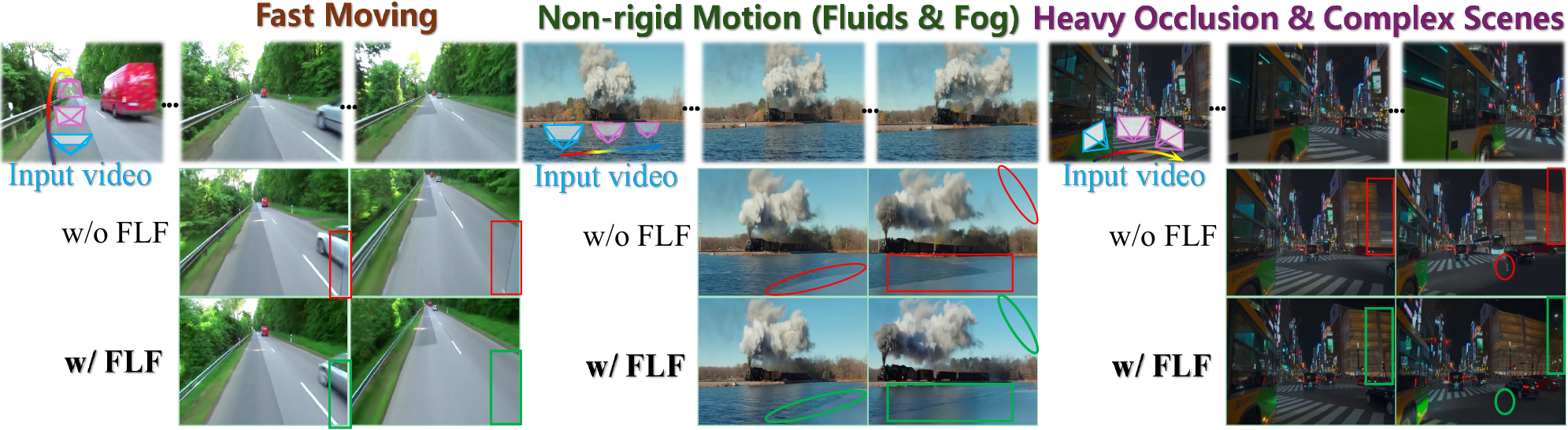}
    \caption{Robustness in challenging scenarios. Our framework maintains structural integrity even under fast motion and complex occlusions.} 
    \label{fig:challenging_scenarios}
    \vspace{-0.5em}
\end{figure*}
\begin{figure*}[htbp]
    \centering
    \includegraphics[width=0.9\linewidth]{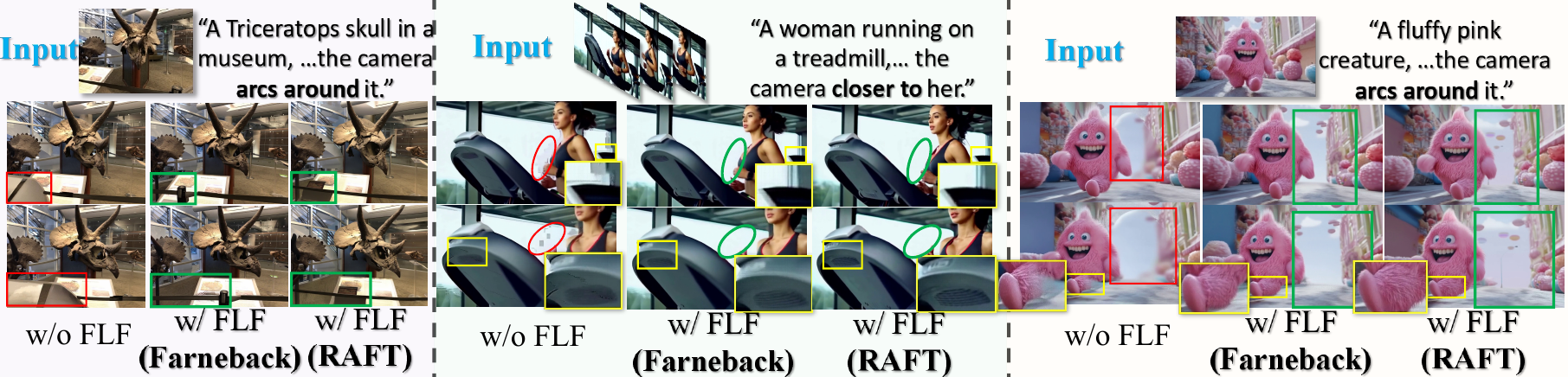}
    \caption{Robustness across optical flow estimators. FLF consistently enhances quality with both Farneb\"ack~\citep{OpticaFlow} and RAFT, validating its flexibility and robustness.} %
    \label{fig:flow_ablation}
    \vspace{-0.5em}
\end{figure*}

\noindent\textbf{Human-Centric Scenes.} Human-centric scenes are challenging for novel view synthesis due to the need for high structural and temporal consistency. As shown in Fig.~\ref{fig:man}, some methods can struggle with these cases, sometimes introducing artifacts, unintended motion, or difficulty rendering plausible facial features. For instance, TrajectoryCrafter~\citep{TrajCrafter} may recover the coarse structure, but can introduce unnatural facial deformations. In contrast, our method's use of strong generative priors and precise trajectory guidance helps maintain scene stationarity and consistency, producing more natural renderings that better preserve the subject's appearance.

\begin{figure*}[htbp]
    \centering
    \includegraphics[width=0.9\linewidth]{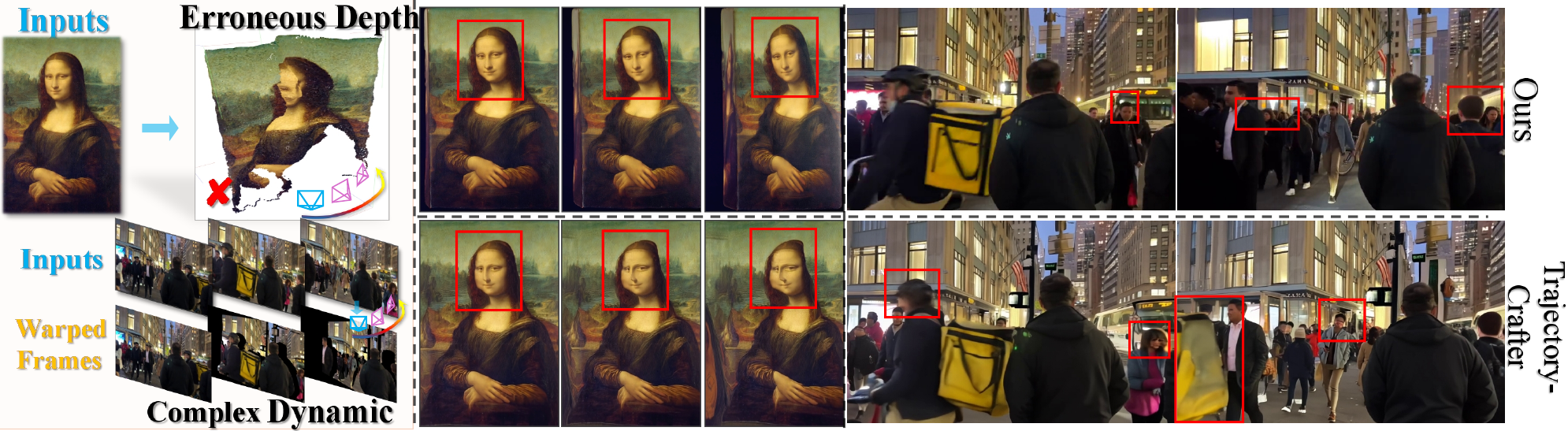}
    \caption{Failure cases. Erroneous depth estimation in highly complex scenes can diminish the control accuracy, leading to artifacts.} 
    \label{fig:failure_cases}
\end{figure*}
\noindent\textbf{Large Camera Movements and 360$^\circ$ View Generation.}
Generating large camera movements (e.g., a 180$^\circ$ turn) or full 360$^\circ$ orbit views from a single image in a single pass is highly challenging for existing methods. It risks hallucination in invisible regions due to the limited field-of-view of the source observation. Our method effectively resolves this problem via an iterative multi-clip generation strategy. By using the last frame of the previous clip as the prior for the next, we successfully achieve large-range scene generation, such as 180$^\circ$ turns (as shown in Fig.~\ref{fig_extreme_motion}). Furthermore, combined with our framework's precise trajectory control, it enables the creation of coherent, object-centric orbit views of complex scenes (Fig.~\ref{fig:360}). We achieve the full 360$^\circ$ loop by generating a sequence where the final frame seamlessly connects to the first. This is made possible by our precise guidance, which maintains high image quality and prevents the accumulation of errors over the entire long-range trajectory, thereby avoiding a common point of failure in other methods. Unlike traditional panoramic approaches, our method directly generates a continuous view along a given trajectory, offering more flexibility and strong visual quality, particularly for object-centric paths.

\noindent\textbf{Challenging Scenarios.} 
We also demonstrate the robustness against fast motion, occlusions, and non-rigid dynamics.
As illustrated in Fig.~\ref{fig:challenging_scenarios}, even when local flow precision drops, enabling FLF consistently yields better results than disabling it. 
This robustness relies on our adaptive designs: when flow is unstable (e.g., high variance), our dynamic threshold automatically adjusts.
This allows more channels to receive control signals, preventing misclassification and maintaining structural stability. 

\subsection{Robustness across Optical Flow Estimators}
\label{Appendix_robustness}
A potential concern is whether our framework heavily relies on a specific optical flow algorithm.
To validate this, we replaced the lightweight Farneb\"ack algorithm~\citep{OpticaFlow} with the learning-based RAFT model. 
As shown in Fig.~\ref{fig:flow_ablation}, our Flow-Gated Latent Fusion (FLF) consistently enhances generation quality regardless of the flow backbone. %
Furthermore, our multi-metric scoring system (evaluating magnitude, direction, and reliability) mitigates single-metric noise, ensuring that the lightweight Farneb\"ack choice is highly sufficient for our pipeline. %

\subsection{Limitations and Failure Cases}
\label{Appendix_limitations}
While our framework achieves precise zero-shot camera control, we acknowledge that, similar to other depth-warping-based methods (e.g., TrajectoryCrafter~\citep{TrajCrafter}), our performance is inherently bottlenecked by the quality of the underlying depth estimation. 

As illustrated in Fig.~\ref{fig:failure_cases}, in scenarios involving extremely complex scene dynamics or severe depth errors, structural distortions may occur.
However, it is worth noting that our dynamic gating mechanism automatically reduces the number of filtered channels in such chaotic cases, ensuring that the outputs are typically no worse than the baseline model without FLF guidance.
In future work, introducing explicit camera pose encoding or semantic priors could help resolve these fundamental ambiguities and further enhance robustness against depth failures.

\subsection{More Cases}
\begin{figure*}[!t]
    \centering
    \includegraphics[width=1.0\linewidth]{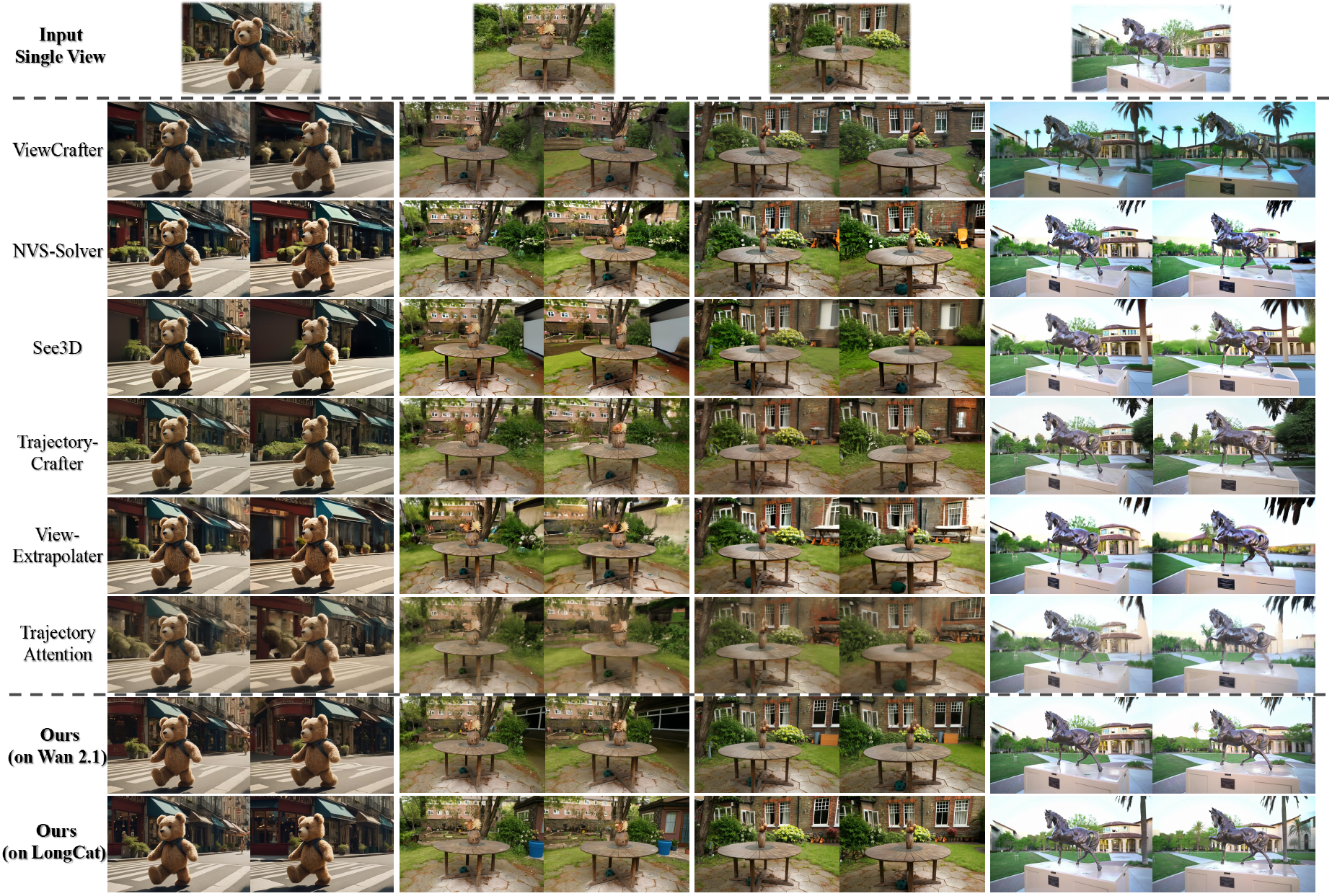}
    \caption{Additional qualitative results for single-view 3D scene generation (Case 1). Validated on Wan 2.1 and LongCat architectures, our method consistently produces 3D-consistent novel views with high visual fidelity.}
    \label{fig:more_3d_1}
\end{figure*}

\begin{figure*}[!t]
    \centering
    \includegraphics[width=1.0\linewidth]{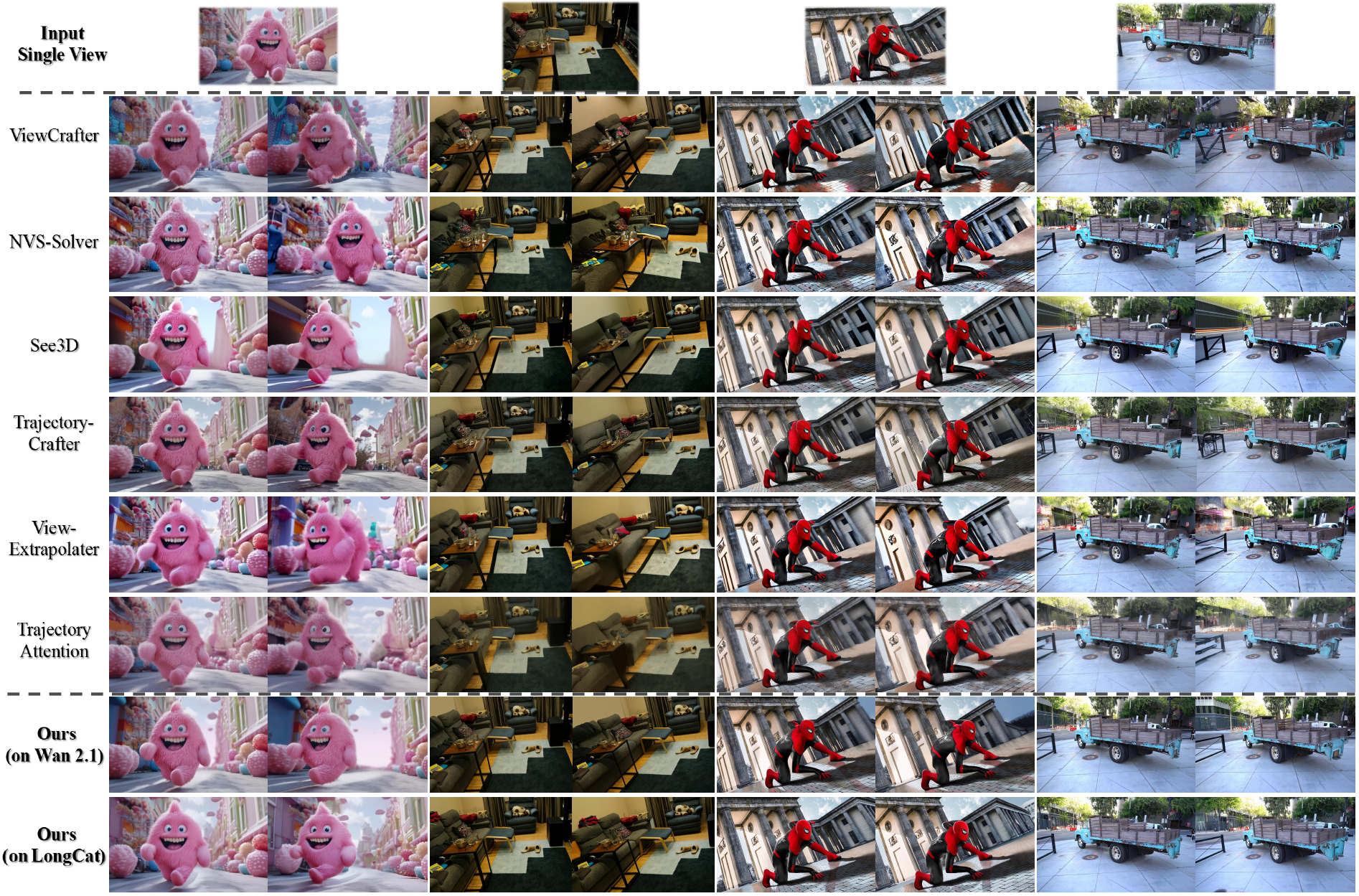}
    \caption{Additional qualitative results for single-view 3D scene generation (Case 2). Validated on Wan 2.1 and LongCat architectures, our method consistently produces 3D-consistent novel views with high visual fidelity.}
    \label{fig:more_3d_2}
\end{figure*}

\begin{figure*}[!t]
    \centering
    \includegraphics[width=1.0\linewidth]{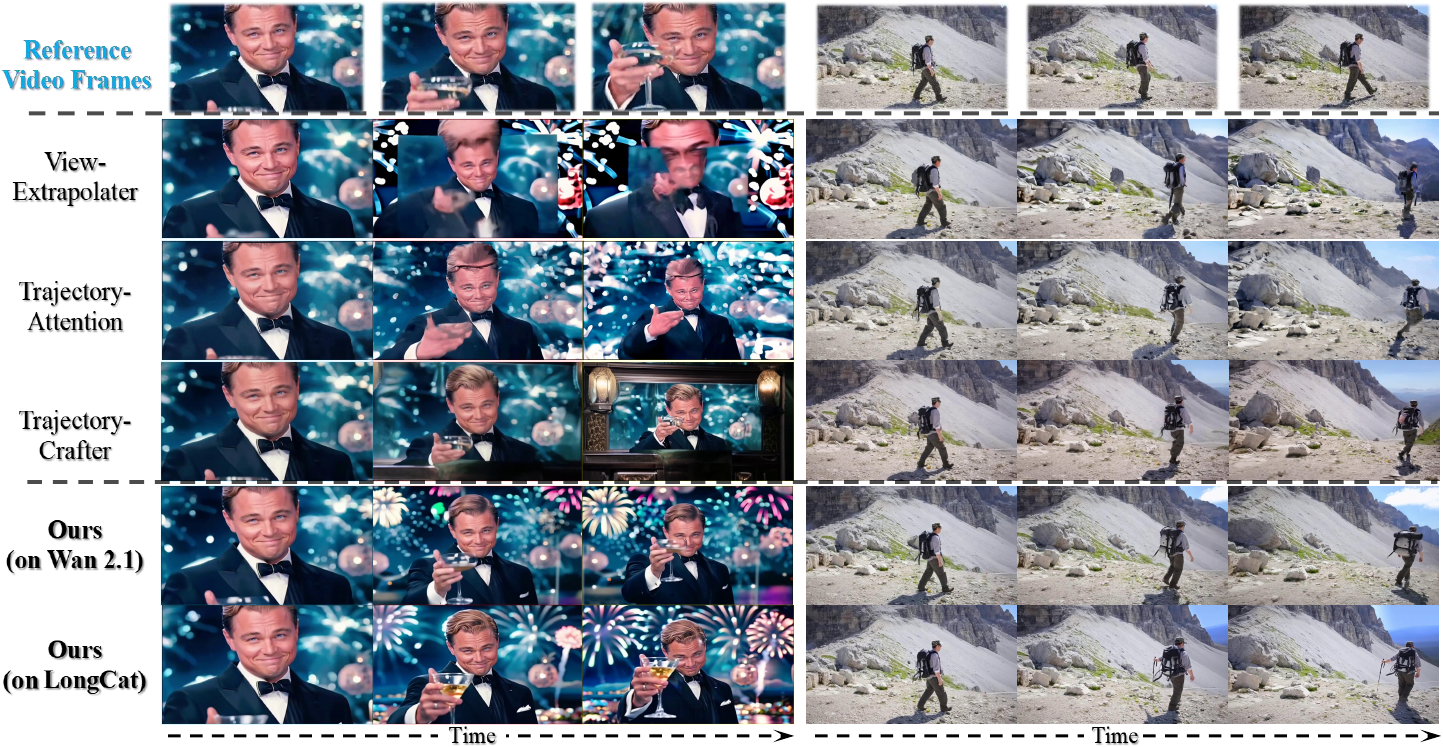}
    \caption{Additional qualitative results for dynamic video re-filming (Case 1). Validated on Wan 2.1 and LongCat architectures, our method enables effective camera control with superior realism and temporal smoothness.}
    \label{fig:more_4d_1}
\end{figure*}

\begin{figure*}[!t]
    \centering
    \includegraphics[width=1.0\linewidth]{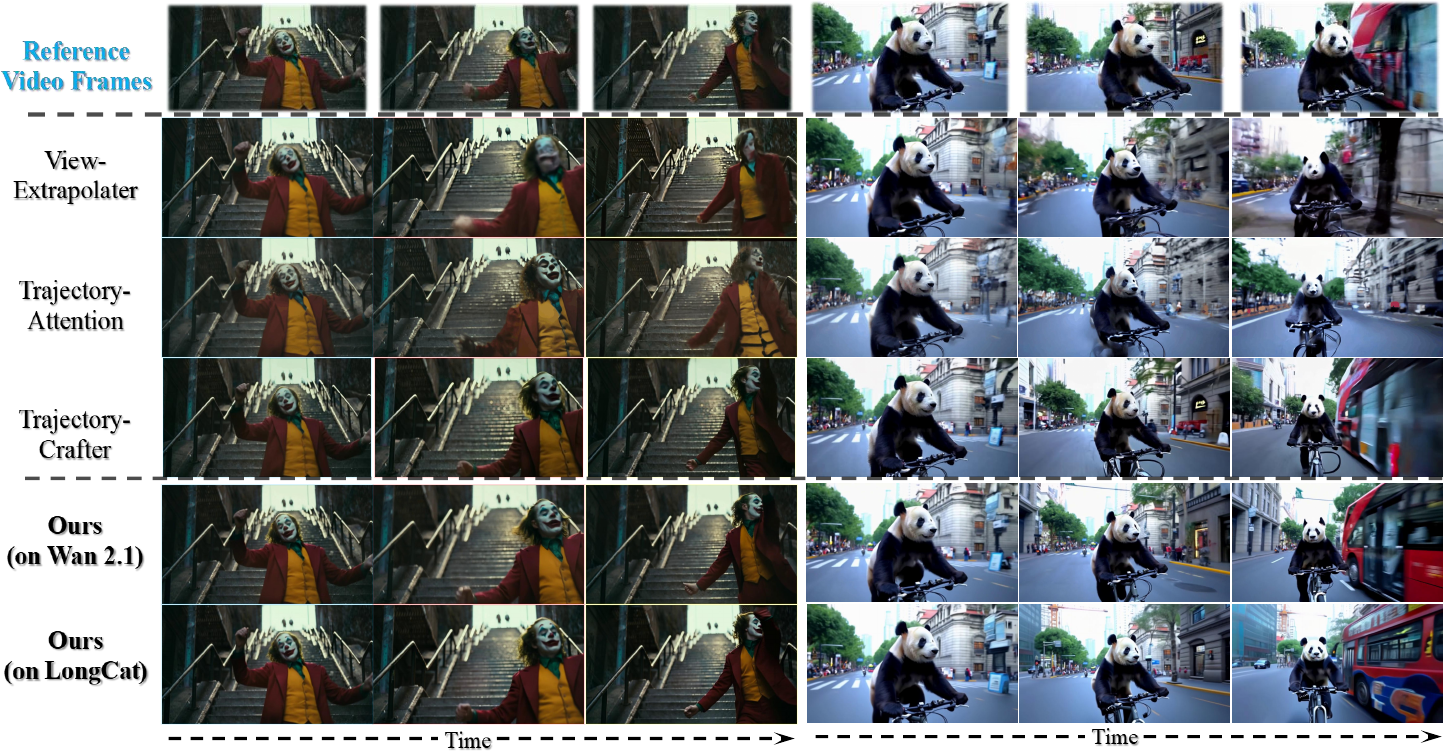}
    \caption{Additional qualitative results for dynamic video re-filming (Case 2). Validated on Wan 2.1 and LongCat architectures, our method enables effective camera control with superior realism and temporal smoothness.}
    \label{fig:more_4d_2}
\end{figure*}

\begin{figure*}[!t]
    \centering
    \includegraphics[width=1.0\linewidth]{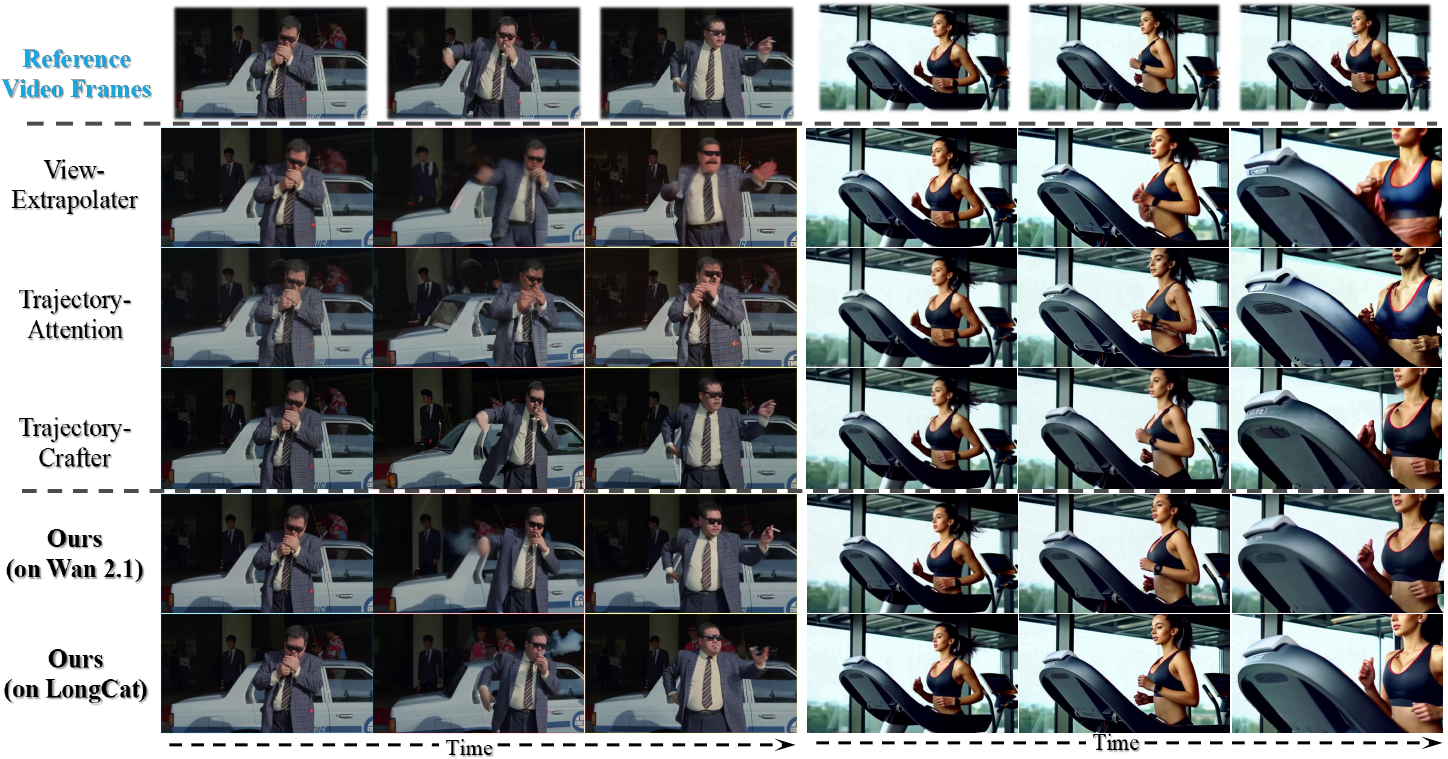}
    \caption{Additional qualitative results for dynamic video re-filming (Case 3). Validated on Wan 2.1 and LongCat architectures, our method enables effective camera control with superior realism and temporal smoothness.}
    \label{fig:more_4d_3}
\end{figure*}

To provide a more comprehensive evaluation of our method's performance across different backbone architectures, we present additional qualitative results in Fig.~\ref{fig:more_3d_1}, Fig.~\ref{fig:more_3d_2}, Fig.~\ref{fig:more_4d_1}, Fig.~\ref{fig:more_4d_2} and Fig.~\ref{fig:more_4d_3}. As illustrated, our approach achieves superior visual fidelity and structural plausibility, consistently delivering state-of-the-art performance on both Wan 2.1~\citep{Wan21} and LongCat~\citep{Longcat} models.

\clearpage

\end{document}